\documentclass[epjST]{svjour}
\usepackage[utf8]{inputenc}
\usepackage[english]{babel}
\usepackage{graphicx}

\usepackage{natbib}
\bibpunct[;]{(}{)}{,}{a}{}{;}

\usepackage{xcolor}
\setcitestyle{notesep={ }}

\begin{document}

\title{
A brief history of Florentine physics from the 1920s to the end of the 1960s}
\author{Roberto Casalbuoni\inst{1}\inst{2}\fnmsep\thanks{\email{casalbuoni@fi.infn.it}}, Daniele Dominici\inst{1} \inst{2}\fnmsep\thanks{\email{dominici@fi.infn.it}} \and Massimo Mazzoni\inst{3}\fnmsep\thanks{\email{mmazzoni.astro@hotmail.com}}  
\institute{Department of Physics and Astronomy, University of Florence,Via G. Sansone 1, 50019 Sesto Fiorentino (FI), Italy \and INFN, Sezione di Firenze, Via G. Sansone 1, 50019 Sesto Fiorentino (FI), Italy \and  Fondazione Osservatorio Ximeniano, Via Borgo San Lorenzo 26, 50123 Firenze, Italy}}

%


%
\abstract{The history of the Institute of Physics at the University of Florence is traced from the beginning of the 20th century, with the arrival of Antonio Garbasso as Director (1913), to the 1960s.
Thanks to Garbasso's expertise, not only did the Institute gain new premises on Arcetri hill, where the Astronomical Observatory was already located, but it also formed a brilliant group of young physicists made up of Enrico Fermi,  Franco Rasetti, Enrico Persico, Bruno Rossi, Gilberto Bernardini, Daria Bocciarelli, Lorenzo Emo Capodilista, Giuseppe Occhialini and Giulio Racah, who were engaged in the emerging fields of Quantum Mechanics and Cosmic Rays.  This {\it Arcetri School} disintegrated in the late 1930s for the transfer of its protagonists to chairs in other universities, for the environment created by the fascist regime and, to some extent, for the racial laws.
After the war, the legacy was taken up by some students of this school who formed research groups in the field of nuclear physics and elementary particle physics. As far as theoretical physics was concerned, after the Fermi and Persico periods these studies enjoyed a new expansion towards the end of the 1950s, with the arrival of Giacomo Morpurgo and above all, that of Raoul Gatto, who created the first real Italian school of Theoretical Physics at Arcetri.
} 
\maketitle

\keywords{Arcetri, Fermi, Garbasso, Rossi, University of Florence}

\section{Introduction}
In 1472, Lorenzo de' Medici decided to transfer the {\it Studium generale et Universitas scholarium}, established in Florence in 1321, to Pisa, due to the great shortage of houses, which would have made it difficult to accommodate students, 
and also because the town of Florence offered too much entertainment to the students.
The Studium never returned to Florence after the move and it was not until 1924 that the University of Florence was established.

Between the actions of Lorenzo de' Medici and 1924, there were intermediate stages that led to the creation of university-type institutions. In 1807, the Queen of the Kingdom of Etruria, Maria Luisa of Bourbon, dedicated the Royal Museum of Physics and Natural History in Via Romana {\it (La Specola)} to education, establishing the Lyceum with six chairs: Astronomy, Theoretical-Experimental Physics, Chemistry, Comparative Anatomy, Mineralogy-Zoology and Botany, with the aim of offering a high-level scientific education, with free programmes and no exams, attendance requirements or enrolment. This organisation of the Lyceum reflected that of the Coll\`ege de France. In turn, the founding principles of this school would be an inspiration for the 1859 Istituto di Studi Superiori Pratici e di Perfezionamento: in fact, the activities of the Lyceum were to produce important discoveries relating to Science and its application for the benefit of the arts and crafts. 
In 1859, the provisional government of Tuscany, which took office after the departure of the Lorraine family, created an institute under this name as an ideal continuation of the Studium generale et Universitas scholarium: an institute without formal enrolment or dissertations, to initiate young people into post-graduate studies. The institute had four sections, one medical, one scientific, one philological and one philosophical. In 1876, it was aligned with the other universities of the Kingdom in terms of its internal organisation. The chair of physics was often vacant until the appointment of Antonio R\`oiti in 1880. It is worth noting that in 1862-63 the chair had been assigned to Pietro Blaserna, called shortly afterwards to the University of Palermo and later, in 1873, to Rome to direct the Institute of Physics. R\`oiti had graduated in Mathematics from Pisa in 1869 under the direction of acclaimed mathematician Enrico Betti. In actual fact, his real teacher was Riccardo Felici, who always sent his students to graduate with Betti because he did not want them in his laboratory \citep{iur2019}. 
	
	After his call to Florence, R\`oiti's research activity turned mainly to electrical measurements, particularly the study of the unit of electrical resistance (Ohm), established for the first time in 1864 by the British Association for the Advancement of Science by a commission chaired by Maxwell. He also held political posts, as town councillor in Florence from 1888 to 1890, and managerial posts, as Dean of the Section of Physical and Natural Sciences from 1894 to 1908.
	
\begin{figure}[h]
\centering
\includegraphics[width=0.7\textwidth]{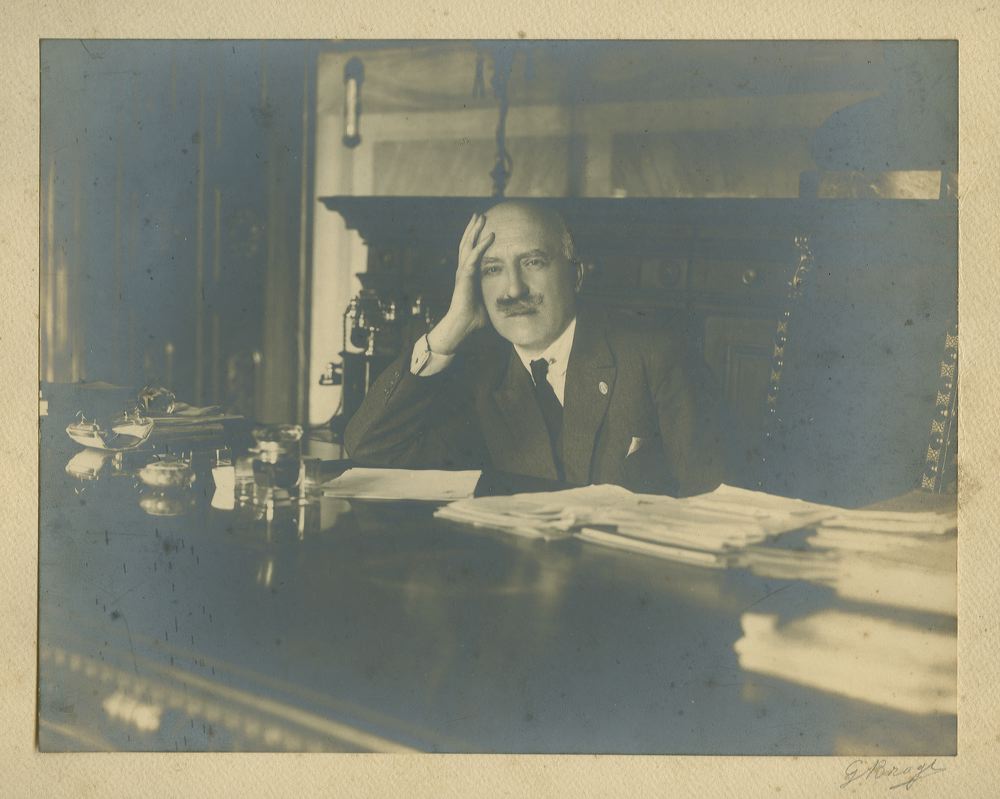}
\caption{Antonio Garbasso, mayor of Florence, in his study (1920) (Fondo Garbasso, University of Florence Historical Archive).}\label{fig1}
\end{figure}

In 1913, after R\`oiti retired, Antonio Garbasso was appointed to the chair of physics. After graduating from Turin in 1892 and having spent time in Germany with famous physicists such as Hertz and Helmoltz, and having held teaching positions in Turin and Pisa, Garbasso was called to the chair of experimental physics in Genoa in 1903, where he remained until he was called to Florence. He worked on optics (explaining the phenomenon of the mirage) and spectroscopy. Garbasso was also an important public figure: he was elected mayor of Florence in 1920  (Fig. \ref{fig1}), standing with the National Bloc, made up of liberals, radicals, republicans and reformists. In 1923, he joined the Fascist Party. He was mayor until 1927, with a short interval of three months, when the city was administered by the Prefectural Commissioner. In 1927, after the dissolution of municipal administrations, he became the first Podest\`a of Florence, replaced in 1928 by Giuseppe della Gherardesca, a nobleman who had participated in gang actions with the Fascists. In 1924, during the crisis that followed Giacomo Matteotti's assassination, Garbasso played an important role in the stability of the Mussolini government by convincing liberal aldermen and town councillors to renew their trust in the government. Mussolini responded with an appreciative telegram. As administrator, he succeeded in improving the city's financial situation by reducing investments in the building, education and health sectors 
 \citep{palla}.

 In the field of physics, he was President of the SIF (Italian Physical Society) for two periods (1912-1914, 1921-1925) as well as of the Committee of Astronomy, Mathematics and Physics of the CNR (National Research Council). He was a delegate of the Ministry of National Education in the Technical Committee for the Optical Industry and played an important role in the political and cultural debate that accompanied Giovanni Gentile's educational reform, opposing the purely humanistic approach to the detriment of scientific disciplines.

As we will see in the next chapter, Garbasso, who combined the characteristics of both scientist and humanist, was a fundamental figure in Florentine physics and can rightly be considered its founding father  \citep{mandstoria,bm2006,bm2007,cdm2021}.

\section{Arcetri school}

\subsection{The first years of the Physics Laboratory in Arcetri, 1913 to 1926}

The period between 1920 and 1940 was a very important one for Italian physics, which achieved international acclaim thanks to the work of a group of brilliant researchers promoted and coordinated by Orso Mario Corbino at the University of Rome and by Antonio Garbasso at the University of Florence.  Orso Mario Corbino became Professor of Experimental Physics at the Royal University of Messina in 1905; in 1918 he succeeded Pietro Blaserna at the Royal University of Rome, taking the chair of Experimental Physics and directing the Institute in Via Panisperna.

Corbino and Garbasso were both excellent physicists. Corbino discovered the Corbino effect, a variant of the Hall effect, and the Macaluso-Corbino effect, a strong rotation, induced by a magnetic field, of the plane of polarisation of light observed at wavelengths close to the absorption line of the material in which the light propagates. As for Garbasso, he tried to give a semiclassical interpretation of the structure of the atom and formulated the theory to explain the Stark-Lo Surdo effect. Apart from this, both men were also interested in the dissemination and politics of science and its practical applications. Corbino was a senator from 1920, Minister of Education (1921-22) and Minister of the National Economy (1923-24), and a member of various boards of directors of electricity companies. As mentioned above, Garbasso was mayor, senator and held other public offices. Both had a positive attitude towards the new concept of Quantum Mechanics and succeeded in building two schools that emerged on the international physics scene: the Via Panisperna boys' school, centred around Enrico Fermi, and the Arcetri school, centred around Bruno Rossi.

\begin{figure}[h]
\centering
\includegraphics[width=0.7\textwidth]{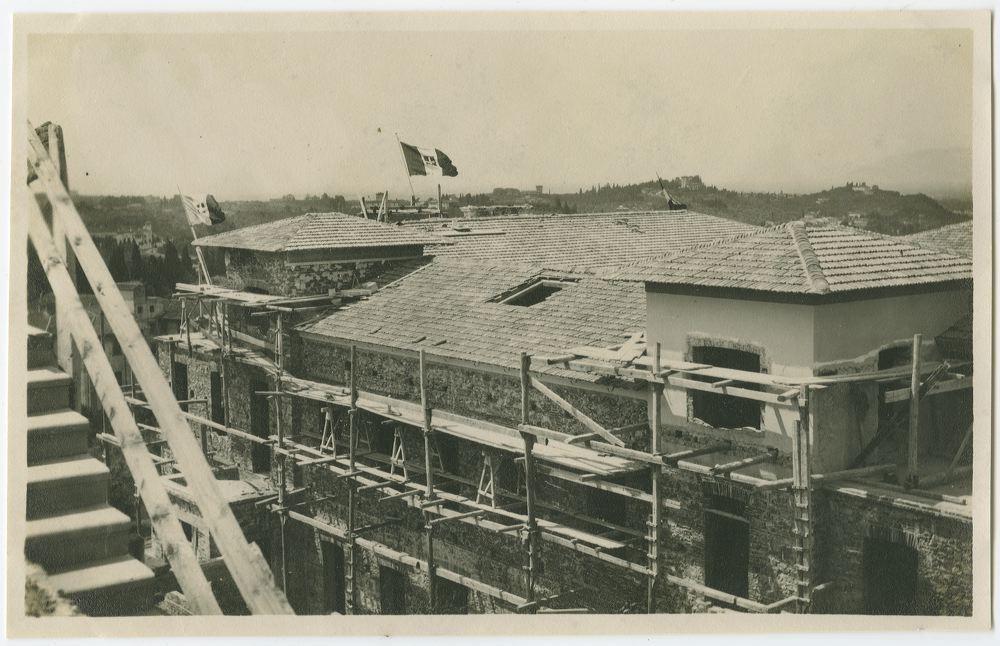}
\caption{The Physics Institute in Arcetri during construction, 1915 (Fondo Garbasso, University of Florence Historical Archive).}\label{fig2}
\end{figure}

	When Garbasso arrived in Florence in 1913, the city still had no university, only the Istituto di Studi Superiori Pratici e di Perfezionamento, which became a formal university in 1924, also thanks to the work of Garbasso.  This created new courses and consequently the possibility, as we shall see, to attract bright young people. Two years after his arrival in Florence, Garbasso succeeded in obtaining funding and permission from the town council to move the Physics Laboratory from the premises in the centre of Florence to a new building on Arcetri hill which already was housing the Astronomical Observatory and was not far from Villa Il Gioiello, where Galileo had lived from 1631 until his death in 1642. And Garbasso, as written by Rita Brunetti, succeeded in {\it giving Florence the prettiest Italian research laboratory with the most remarkable location }\citep{brunetti1933}.
	On 24 June 1916 a ceremony was held in Arcetri to celebrate the completion of the roof on the building, the construction of which had begun in 1915 (Fig.~\ref{fig2}). The project also included the construction of a pavilion for Physics of the Earth for the activities of Antonino Lo Surdo, who had been Director of the Via Romana Meteorological Observatory since 1910. But Lo Surdo was called to Rome in 1918 and the building remained unused for several years. 

Garbasso was also very interested in technological outcome and became involved in this area, inaugurating the Laboratory of Practical Optics and Precision Mechanics in 1918 in Arcetri, in the premises of the pavilion of Physics of the Earth, which was first affiliated to the University of Florence and later became the National Institute of Optics \citep{brunetti1933,abetti1933}.
 
\begin{figure}[h]
\centering
\includegraphics[width=0.7\textwidth]{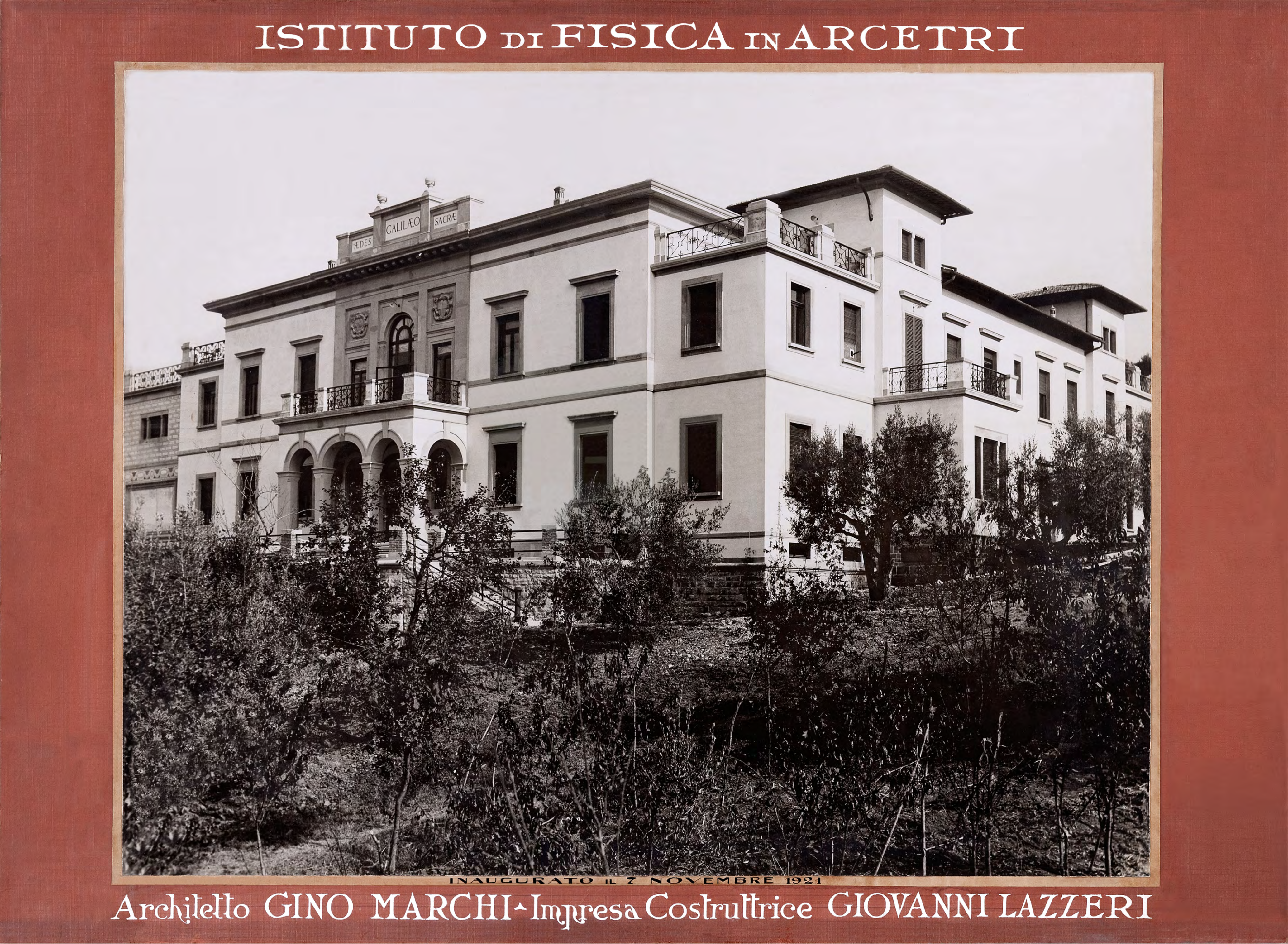}
\caption{Arcetri Physics Institute in an old photograph (Department of Physics and Astronomy, Scientific and Technological Hub, University of Florence).}\label{fig3}
\end{figure}

 The new Arcetri site of the Physics Laboratory, the name that had been given to the old Cabinet of Physics in 1918, was inaugurated in 1921 (Fig.~\ref{fig3}). As proof of the esteem in which the scientific institutions on the hill were held, they were visited by King Vittorio Emanuele in 1921 and by Mussolini in 1923  \citep{abetti1933}.

	In addition to the construction of the new building, it was essential to start recruiting bright young people. Garbasso's work began in 1922, when he hired Franco Rasetti, who had just graduated from Pisa with a thesis on spectroscopy with Luigi Puccianti, who was head of Experimental Physics, as assistant for Physics of the Earth. At that time, Antonio Garbasso's assistants were Rita Brunetti, the first woman in Italy to win a chair in Physics, in Ferrara in 1926, and later, from 1928, to hold the position of Director of the Institute in Cagliari, and Vasco Ronchi, who became Director of the National Institute of Optics when it was founded in 1930.
	
On the subject of the Florentine period, Rasetti's recollections are very interesting: {\it My first job was in Florence, and I worked on atomic spectroscopy. The physics
building was on a hill near where Galileo lived the last years of his life - in the Arcetri hills.
The physics building had been built there and was extremely inconvenient for students, because
all the other courses were given in various places in town and this was three kilometers out of
town, and also at least 150 meters above the city level. So one had to take a streetcar, and after
that it was still a fairly long walk to get up there. And on this Arcetri hill, the university owned a
fairly large estate, on which the physics building was built, very beautiful architecturally but
very impractical - the most impractical place because it was impossible to heat. It was built like
an abbey: it was a rectangular building with a vast garden and lawn in the middle}  \citep{goods}.

	Rasetti again, in relation to the laboratories:  {\it The equipment was pretty good for those times - especially
for spectroscopy, which was my field. They had a very good spectrograph and spectroscope; we
had an excellent Rowland grating in the Rowland mounting. And I didn't have much teaching to
do, because [Antonio] Garbasso gave the physics course} \citep{goods}.

On the subject of Garbasso, Rasetti says: {\it Garbasso had been a good physicist, but when I knew him he was only interested in politics. He was the mayor of Florence} [from 1920]. {\it  He gave his course in elementary physics and he was
quite intelligent at it. And later Fermi explained to him what we were doing and he understood,
because he was intelligent. I mean, he knew the classical theory - he didn't know much about
 the quantum theory, because that had come after he lost direct interest in physics. But he
followed what we were doing, and he was a very pleasant person} \citep{goods}. 

	Unfortunately, the Institute wasn't in the best financial situation. As Bruno  Rossi remembers:  {\it The institute was always late paying its electricity bill, and the only reason we were not cut off was that the director was (or had been until recently) the mayor of the city}  \citep{rossibook}. 
	
	In academic year 1924/25 Garbasso offered a position as lecturer to the young Enrico Fermi, to cover courses in Theoretical Mechanics and Mathematical Physics, which was renewed the following year.

After graduating from the Scuola Normale Superiore in Pisa, Enrico Fermi spent two periods abroad, in G\"ottingen, Germany, and Leiden, Holland, with an interlude in academic year 1923/24 when he took up a temporary position in Rome, teaching Mathematics for Chemists. Corbino, whom Fermi had met in 1922, had helped him to obtain both the scholarship to G\"ottingen and the temporary assignment. But thanks to some contacts that Fermi had made in Florence, Garbasso was the first to offer Fermi a teaching position in Physics.

The course in Teoretical Mechanics was taken by students in the undergraduate courses in Physics, in Physics and Mathematics, in Mathematics and in the two-year  engineering preparatory course. Two of the students  for engineering studies in academic year 1925-26 reorganised the notes of the Teoretical Mechanics course into handouts that were printed in 1926 by Litografia Tassini in Florence\footnote{A book containing these lessons was recently published by Firenze University Press \citep{cdp2019}.}. 
	
	The course in Mathematical Physics, taught in the fourth year of the degree courses in Physics, Physics and Mathematics and in Mathematics, included for 1924/25 the traditional topics of Electrodynamics supplemented by a mention of the new Theory of Relativity.
	
In the following year, the title of the course was changed to Theoretical Physics and Fermi covered notions of Probability, Thermodynamics and Statistical Mechanics.  This course in Theoretical Physics in the academic year 1925/26, together with a similar course held in Naples, were the first with this title in Italy\footnote{For a reconstruction of the institution of the first chairs of Theoretical Physics in Italy, see   \citep{larana1}. }.  

It was in those years, therefore, that the teaching of Theoretical Physics was born in Italy and it was to this chair in Florence that Enrico Persico was appointed the following year. In other European universities some chairs of Theoretical Physics had already existed since the middle of the 19th century, albeit in small numbers. In 1900, the number of chairs in Theoretical Physics worldwide was eleven, eight in Germany, two in the United States and one in the Netherlands 
\citep{pais1985}. 

Enrico Fermi's time in Florence was short but very fruitful, thanks in part to the presence of his old Scuola Normale Superiore classmate, Franco Rasetti. Fermi kept Rasetti up to date on quantum mechanics and Field Theory. In turn, Rasetti taught Fermi the art of experimenting with spectroscopic techniques, of which he was a real master. Fermi was one of the last scientists versed in theory and experiments as he demonstrated in the Roman period with his studies on induced radioactivity. During that time, Fermi and Rasetti wrote several articles together. Franco Rasetti would later speak of "Fermi's second venture in the experimental field after several years of theoretical works" \citep{fermioc}. One of these experimental works proved to be of some significance: the two friends analysed the effect of weak but high-frequency magnetic fields on the depolarisation of resonance light in mercury vapour. This work was the first example of the study of atomic spectra using radio-frequency fields, a technique that would receive numerous applications in the years to come \citep{rasfermi1925, rasfermi1925a}. 

His wife, Laura Capon, describes the friendship between Fermi and Rasetti in her book, Atoms in the Family
\citep{caponl}: {\it The physics laboratories of the University of Florence were in Arcetri, on the famous hill where Galileo had lived during the last years of his life and where he had died. Accompanied by his friend Rasetti, Fermi spent long hours hunting geckos, small, completely harmless lizards. Fermi and Rasetti would then release the geckos they had caught into the dining room for the pleasure of frightening the girls serving at the tables. The two friends would spend hours lying on their stomachs in the grass, perfectly still, holding a glass rod with a small silk noose at the other end. During this vigilant wait Rasetti would observe the small world before his eyes, a tender leaf of grass, a busy ant carrying a piece of straw, the effect of the sun reflected on the glass rod} \citep{caponl}. Rasetti had many passions and later became an expert in geology, palaeontology, entomology and botany. During the Canadian period in particular, he became famous for his study of the Cambrian trilobites.

	During his Florentine period, Fermi stayed in the so-called {\it vagoncino}, the premises of the Physics of the Earth, which later became the first headquarters of the National Institute of Optics in Arcetri. This building had a room with a bed and a stove that had allowed Rasetti to stay in the company of scorpions for the previous two years \citep{goods}. After the death of Rasetti's father, his mother moved to Florence and so Rasetti left the vagoncino to live with his mother.
	
	In 1926, Fermi published the work \citep{fermistat1,fermistat2}, that made him internationally famous, in which he deduced what is now called the Fermi-Dirac statistics and from which the name fermions originated. Since 1923, Fermi had been interested in Statistical Mechanics and particularly in the problem of the absolute constant of the entropy of a perfect gas, i.e. the Sackur-Tetrode formula  \citep{cordella2000,cordella2001}. The new factor, which enabled him to reach the Fermi-Dirac statistics, was the Exclusion Principle formulated by Pauli in 1925. This was Fermi's great merit, having applied Pauli's Principle, which until then had been advanced for the interpretation of spectroscopic phenomena, to a general system of particles. It is also interesting to report Cordella and Sebastiani's observation that Fermi may have found the starting point for the reflections, which led him to the new quantum statistics, while preparing the lectures on Statistical Mechanics, which he was to give in Florence in 1925/26.

According to  Bruno Pontecorvo  {\it Fermi had been entertaining the idea of this work for some time, but he lacked the Pauli Principle.  As soon as this was formulated, he had his article printed.  It must be said that Fermi was bitterly disappointed at not having been able to formulate the Pauli Principle on his own, despite having come very close, as can be seen from his work} \citep{pontecorvo1993}. 
	
	Paul Dirac also arrived at the same result as the Roman scientist in August 1926. After the publication of Fermi's work, important applications by \citep{thomas1927}, independently of Fermi \citep{fermi1927}, treating the internal electrons of a heavy atom with statistics, and by Fowler \citep{fowler1926}, on white dwarfs and Sommerfeld \citep{sommerfeld1927} on conduction in metals, followed.
	
In 1925 and 1926, Fermi began to worry about his academic future, as shown by his correspondence with Persico. In 1925, he was awarded the "Libera Docenza" in Mathematical Physics and attempted his first selection for a chair, that of Mathematical Physics, at the University of Cagliari. The position was awarded by a majority vote to Giovanni Giorgi, inventor of the MKS System. Fermi was supported by the two mathematical physicists on the Commission, Tullio Levi Civita and Vito Volterra, who were more aware than the other commissioners of the importance of new 20th century physics. 

With the support of Roman mathematicians Guido Castelnuovo, Federigo Enriques and Tullio Levi Civita, in 1926 Corbino succeeded in having the first chair of Theoretical Physics  advertised at the Roman university. Following the selection process, Fermi was called to Rome, Persico to Florence and Pontremoli to Milan.

The arrival of Persico  in Florence on the chair of Theoretical Physics  was of great importance for his extraordinary teaching skills and for his contribution to the spread of Quantum Mechanics. His lectures were collected by Bruno Rossi,  "Aiuto" of Garbasso, called in 1927 from Bologna, and by the student Giulio Racah. These lecture notes, known as the Vangelo Copto,  were the basis for the Persico book on Quantum Mechanics \citep{persico1950}.

 \subsection{The birth and the activities of the Arcetri School from 1927 to 1933}

With the arrival in Florence of Bruno Rossi in 1927 and Gilberto Bernardini in 1928, along with the graduations of Giuseppe "Beppo" Occhialini in 1929, Giulio Racah and Daria Bocciarelli in 1931 and Lorenzo Emo Capodilista in 1932, a group of young people who greatly boosted cosmic ray research was formed\footnote{For Bruno Rossi's contribution to the physics of cosmic rays, see  \citep{Peruzzi:2007zz,Bonolis:2011ph}}. 

These studies had commenced following a famous article in 1929 by Walter Bothe and Werner Kohlh\"orster, which showed that the cosmic radiation observed at sea level was not due to electromagnetic radiation, as thought by Millikan, but actually consisted of ionising particles, and it was assumed that primary radiation (which reaches the atmosphere from sidereal space) was also corpuscular. The technique used was based on the newly invented Geiger counters, which signalled the passage of a single particle with an electrical pulse, and the coincidence technique, i.e., the recording of simultaneous pulses by counters placed one above the other. This made it possible to detect the passage of the same ionising particle through two or more counters in order to detect false signals.

As Rossi reminisced{\it ...and so began one of the most amazing periods of my existence. Was it the thrill of being the first to venture into unknown territory? Was it the special atmosphere created by the relationships between the friends in Arcetri? Was it the subtle charm of the Tuscan hills? }  \citep{rossibook}. Rossi referred to what was later called {\it the spirit of Arcetri} [as remembered  in \citep{bm2007}].

\begin{figure}[h]
\centering
\includegraphics[width=0.7\textwidth]{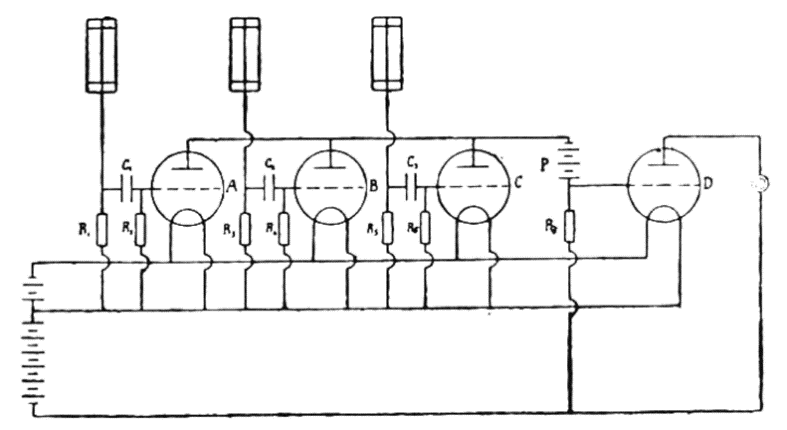}
\caption{Rossi's circuit for detecting cosmic ray coincidences arriving on Geiger counters (The rectangles at the top of the diagram), by \citep{RossiMethodOR}.}\label{fig4}
\end{figure}

\begin{figure}[h]
\centering
\includegraphics[width=0.7\textwidth]{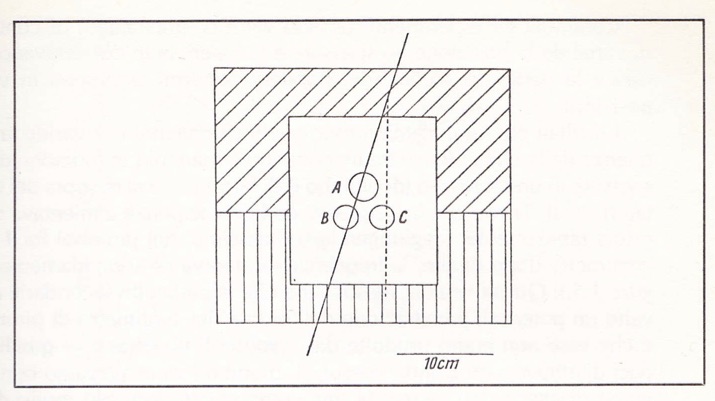}
\caption{The use of the Rossi circuit to detect a triple coincidence which, in the drawing of the three counters, shows the production of secondary radiation (dashed line) by primary radiation (solid line), by \citep{Rossitriple}.}\label{fig5}
\end{figure}

In relation to the research on cosmic rays, Rossi observes: {\it I immediately set to work. The group's solidarity manifested itself in the offer of generous cooperation...} (ibidem). As a result of this work, the famous Rossi circuit, consisting of triodes, was born \citep{RossiMethodOR} (Fig.~\ref{fig4}), allowing the detection of triple coincidences of ionising particles (Fig.~\ref{fig5}) \citep{Rossitriple}.


With this new electronic coincidence apparatus at his disposal, improving the time resolution of Bothe's and Kohlh\"orster's experiment, tenfold to one thousandth of a second, Rossi set to work and performed an experiment with magnetic lenses to measure the charge of the corpuscles that made up the cosmic rays \citep{RossiOnTM,Rossi1931MagneticEO}. The suggestion to use magnetic lenses was made by Luigi Puccianti during a visit by Rossi to Pisa \citep{Rossi1931MagneticEO,RossiOnTM}. Assuming that cosmic rays were electrons, Rossi expected to measure a negative charge.  Surprisingly, he found a slightly positive charge.  The particle Rossi was observing was the lepton $\mu$ or muon, whose mass was measured seven years later. As the mass of $\mu$ was between the mass of the electron and that of the  proton, for a while the particle was known as the {\it mesotron}. And, in fact, there are slightly more positively than negatively charged  $\mu$  at ground level.

In 1931, an important international conference on nuclear physics was held in Rome, attended by numerous Nobel Prize winners and leading physicists of the time.  The scientific secretary of the conference, strongly supported by Corbino, was  Fermi. The only Italian invited speaker   was Bruno Rossi, who gave a talk on cosmic rays. In his talk, Rossi showed that cosmic rays consisted essentially of charged particles and not $\gamma$-rays, as claimed by the American physicist Robert Millikan, Nobel Prize winner in 1923 for his work on the charge of the electron and the photoelectric effect. As stated by Rossi: {\it my speech received a mixed reception. Millikan evidently could not admit that his beloved theory was being ruthlessly attacked by a young man of only 26, so he refused to acknowledge my existence. On the other hand, my speech aroused the interest of Arthur Compton who had never worked on cosmic rays before. Later, he was kind enough to tell me that his interest in cosmic rays arose from my presentation} (ibidem).
This conference was the first international recognition of the young leading school of modern physics in Italy.

In another experiment, Rossi demonstrated that the interaction of cosmic rays with matter can produce showers of secondary particles. During his few years in Florence, Rossi, Bernardini and their young collaborators wrote numerous articles on the absorption of cosmic rays and their behaviour in the earth's magnetic field.

\begin{figure}[h]
\centering
\includegraphics[width=0.7\textwidth]{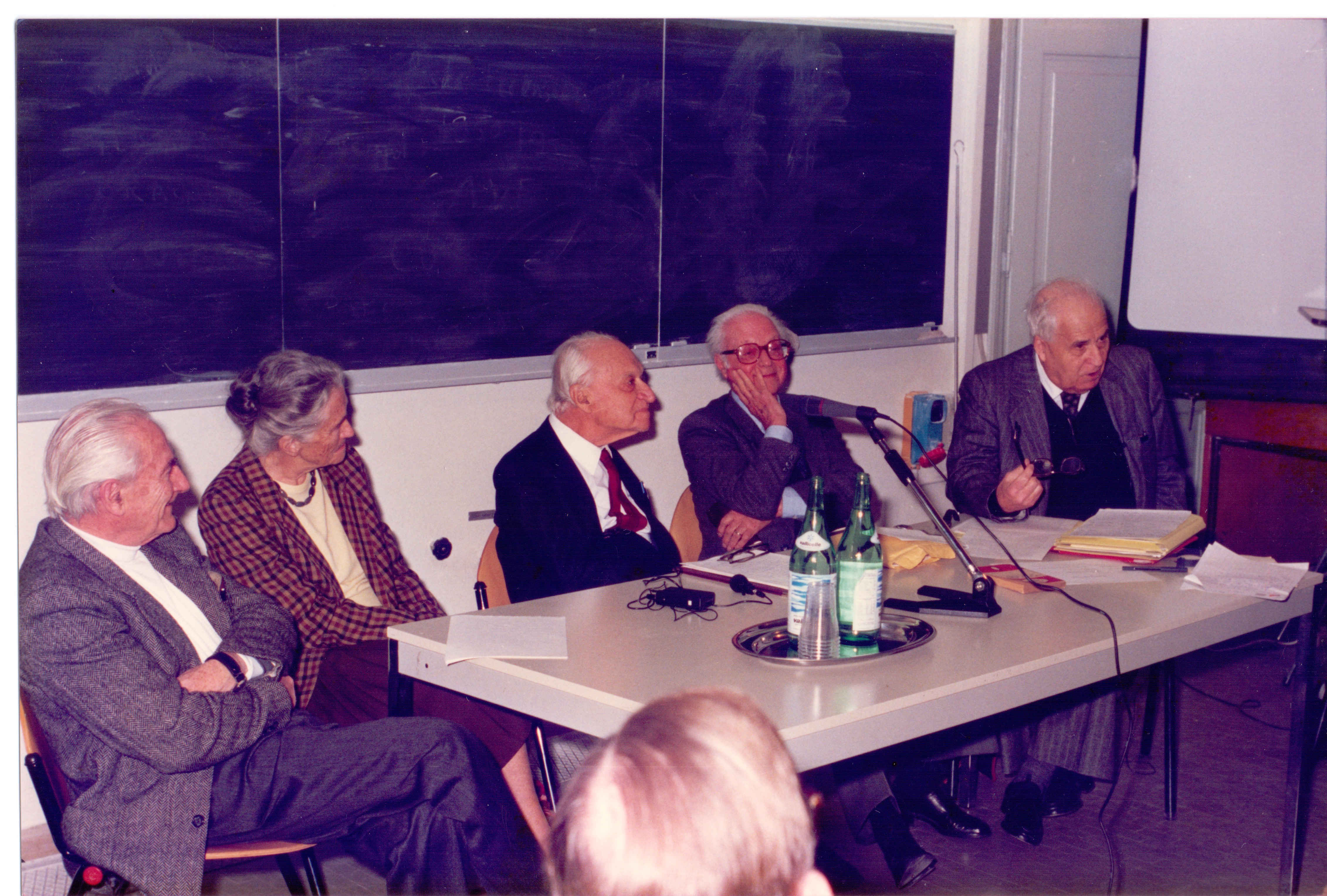}
\caption{Round table at the conference held in Arcetri in 1987. From left: Bernardini, Bocciarelli, Rossi, Amaldi, Mand\`o, (courtesy of Pier Andrea Mand\`o).}\label{fig6}
\end{figure}

This Florentine period was remembered in the conference, known as the conference of the three greats, held in Arcetri in 1987 on Giuseppe Occhialini's 80th birthday (Fig.~\ref{fig6}), attended, among others, by Edoardo Amaldi, Gilberto Bernardini, Daria Bocciarelli, Manlio Mand\`o, Bruno Rossi and Giuseppe Occhialini \citep{bm2007}. On this occasion, the fundamental contribution of Giorgio Abetti, then Director of the nearby Observatory, was remembered. Under his impetus, the Mathematical, Physical and Astrophysical Seminar of Arcetri was launched in 1928. This was a series of international conferences aimed at all students and professors of the university, which was officially approved by the Faculty in 1932 and ran until 1943. This seminar was of great importance for the young people of the Institute of Physics, because it gave them the opportunity to get to know many world famous scientists, who were invited by Abetti on a regular basis. On this subject, Edoardo Amaldi, again at the above-mentioned conference, referring to his attendance at the seminar, reported his impressions of Abetti, describing him as an exceptional figure, capable of taking an interest in any problem of physics and astrophysics. 
It has also to be stressed that it was Abetti that in 1925 built the second solar tower in Europe after the Postdam one. Hale scientific and financial help was instrumental for the design and the realization of Arcetri tower \citep{gmr2004}.

The consequences of studying cosmic ray physics were manifold. A particularly important one stemmed from the lack of expertise on cloud chambers in Italy, which were fundamental instruments for determining the characteristics of particles. Patrick Blackett, in Cambridge, was Europe's leading expert on the subject and Rossi decided to send Occhialini to work with Blackett. Occhialini left in 1931, taking the skills acquired in the field of coincidences at Arcetri with him. The idea was to combine the Rossi circuit with the cloud chamber. Blackett and Occhialini obtained their first results in 1933.  The most exciting result was the discovery of the shower produced by cosmic rays and of the  formation of electron-positron pairs \citep{Bonolis:2011ph,Bonolis:2012dm}. Moreover, thanks to the cloud chamber immersed in a magnetic field, it was possible to observe the components of the shower and also determine the sign of the particle charge. Occhialini's contribution was extremely important for the identification of the positron thanks to Rossi's circuit. This enabled them to confirm Carl Anderson's discovery of the positron (1932), published just a few months earlier.

Unlike the others, Racah, who was a theorist, was not very involved in the research carried out by the cosmic ray group. From 1932 to 1937, he taught a course in theoretical physics, and then moved to Pisa, and in 1939, following the enactment of the racial laws by the Italian government, to Palestine. He often travelled to Rome for talks with Enrico Fermi, Ettore Majorana and Gian Carlo Wick. During his years in Florence, Racah worked on calculations of bremsstrahlung cross sections from high-energy electrons and the production of electron-positron pairs, physical quantities relevant to the study of cosmic rays, and hyperfine structures in atoms. Racah had also established an important working relationship with Wolfgang Pauli in Zurich and published several works with him.

\subsection{The relations between Florence and Rome groups, the birth of nuclear physics in Italy, and the eventual dispersal of the Arcetri group in the late 1930s}

Relations between the Florence and Rome groups in those years were very close.  Rasetti, who had been Garbasso's assistant in 1922, became Orso Mario Corbino's assistant in 1926, filling the position vacated by Persico, who had been called to Florence to the chair of Theoretical Physics.  This exchange of personnel strengthened relations and collaborations between the two groups, who soon achieved international fame working on different topics, cosmic rays in Florence and atomic spectroscopy, the Raman effect and, from 1932, nuclear physics in Rome, as remembered by Edoardo Amaldi   \citep{bm2007}.
	
	During one of these visits on a weekend in 1933, Bruno Rossi, after a discussion with his Roman colleagues on the effect of the earth's magnetic field, wrote an article on this subject in collaboration with Fermi \citep{rossifermi}. Rossi had in fact conjectured the existence of an east-west asymmetry in the distribution of cosmic rays, due to the effect of the earth's magnetic field, which predicted the arrival of more particles from the east or west of the magnetic meridian depending on whether the particle's charge was negative or positive. The experiment in Florence had proved negative. In the work with Fermi, they showed that the explanation for the negative result required that large absorption by the atmosphere and that the east-west effect would be visible near the equator. Rossi, also helped by Garbasso, began to organise a mission to Eritrea to reveal this effect. The mission could only be held in 1933, after Rossi moved to Padua. Sergio De Benedetti and Ivo Ranzi also participated in the mission. The experiment definitively confirmed the corpuscular theory of cosmic rays and that the prevailing direction of the particles was from the west of the magnetic meridian and that the particles were therefore positively charged.

Sergio De Benedetti, born in Florence on 7 August 1912, approached the anti-fascist organisation {\it Giustizia e Libert\`a} as a student in the 1930s.  After graduating in Physics in Florence in 1933, he moved to Padua for a year, to continue his collaboration with Rossi, and in 1935 to Paris. After returning to Italy, he was forced to flee because of the racial laws, first to France, where he participated in the movement to reorganise Italian anti-fascist emigrants, and then to Lisbon in 1940 where he boarded a ship bound for the United States.   He was a professor at the Carnegie Institute of Technology and secretary of the local Federation of Atomic Scientists, and was involved in the nuclear arms control movement.  He died in Florida in 1994.

Thanks to the good relationship between the physics institutes in Florence and Rome, when Fermi needed Geiger counters to study neutron-induced radioactivity, the Florentines supplied all their expertise. Amaldi again recalls \citep{bm2007}  that when Rossi had already moved to Padua, {\it one weekend in April or May 1934, Bernardini, Occhialini, Daria Bocciarelli and Emo Capodilista came to Rome; they brought us boxes full of Geiger counters and proportional counters: they were a gift to help us in our work... They were beautiful and worked very well, but unfortunately the geometry was wrong... }(ibidem). An interesting hypothesis has been put forward to this end \citep{gr2015}, that among the counters tested by Fermi and mentioned in the Roman scientist's notebook, found by Francesco Guerra and Nadia Robotti in Avellino,   there were those  brought by the Florentines. The presence of a large brass counter in this list would suggest cosmic ray counters. According to the two researchers, {\it if, on the other hand, the visit reported by Amaldi actually took place in March 1934, then the entire reconstruction of the counters used in Rome would have to be reconsidered, and the role of Arcetri would be amplified.}

Rossi had, in fact, travelled to Berlin in 1930 on a CNR scholarship, procured by Garbasso, to spend some time in Bothe's laboratory. In Berlin, Rossi realised that Bothe's counters were better than his own, until one day Bothe confessed to him: {\it  I will tell you a secret, but you must promise not to tell anyone about it} \citep{rossibook}.  After Rossi had promised, Bothe continued {\it My counters do not have a steel wire, as people think, but an aluminium wire} (ibidem).  That was how Bothe's secret came to Italy. Or as Rossi remembers: {\it To my shame, I must confess that, back in Italy, I did not feel able to keep the matter of the aluminium wire a secret from my friends in Florence and Rome, but tried to ease my conscience by asking them for the same promise of secrecy that Bothe had requested from me} (ibidem).  In any case, this episode is further confirmation of the close relations between the two groups and of the importance of Rossi's role in bringing Bothe's secret, on the Geiger counter construction technique, from Berlin to Rome via Florence.

The commencement of nuclear research in Italy dates back to 1933 and the institutes involved included not only Rome and Padua, but also Florence. It has to be stressed  that this programme, supported by National Research Council,  assigned the Florentine institute the task of studying the excitation of neutrons in various elements with $\alpha$-particles of different energies, as well as the disintegrations produced by neutrons as they pass through matter\citep{gr2015}. This programme involved Bernardini, Bocciarelli and Capodilista, but was essentially carried out abroad by Bernardini and Emo Capodilista in Berlin Dahlem, where Lise Meitner was also working. The group in Padua was supposed to focus on cosmic rays while the group in Rome was supposed to focus on $\gamma$-spectroscopy. The Rome group, as is well known, under the leadership of Fermi was to discover neutron-induced radioactivity in 1934, paving the way for nuclear fission.

In 1932, Rossi moved to the chair of Experimental Physics of the University of Padua. After his departure the research activity on cosmic rays in Arcetri was carried on by Bernardini, Bocciarelli, Capodilista and Franchetti. The activity of these researchers was also devoted to the study of induced radioactivity, like for example in the polonium-beryllium system. 

\begin{figure}[h]
\centering
\includegraphics[width=0.7\textwidth]{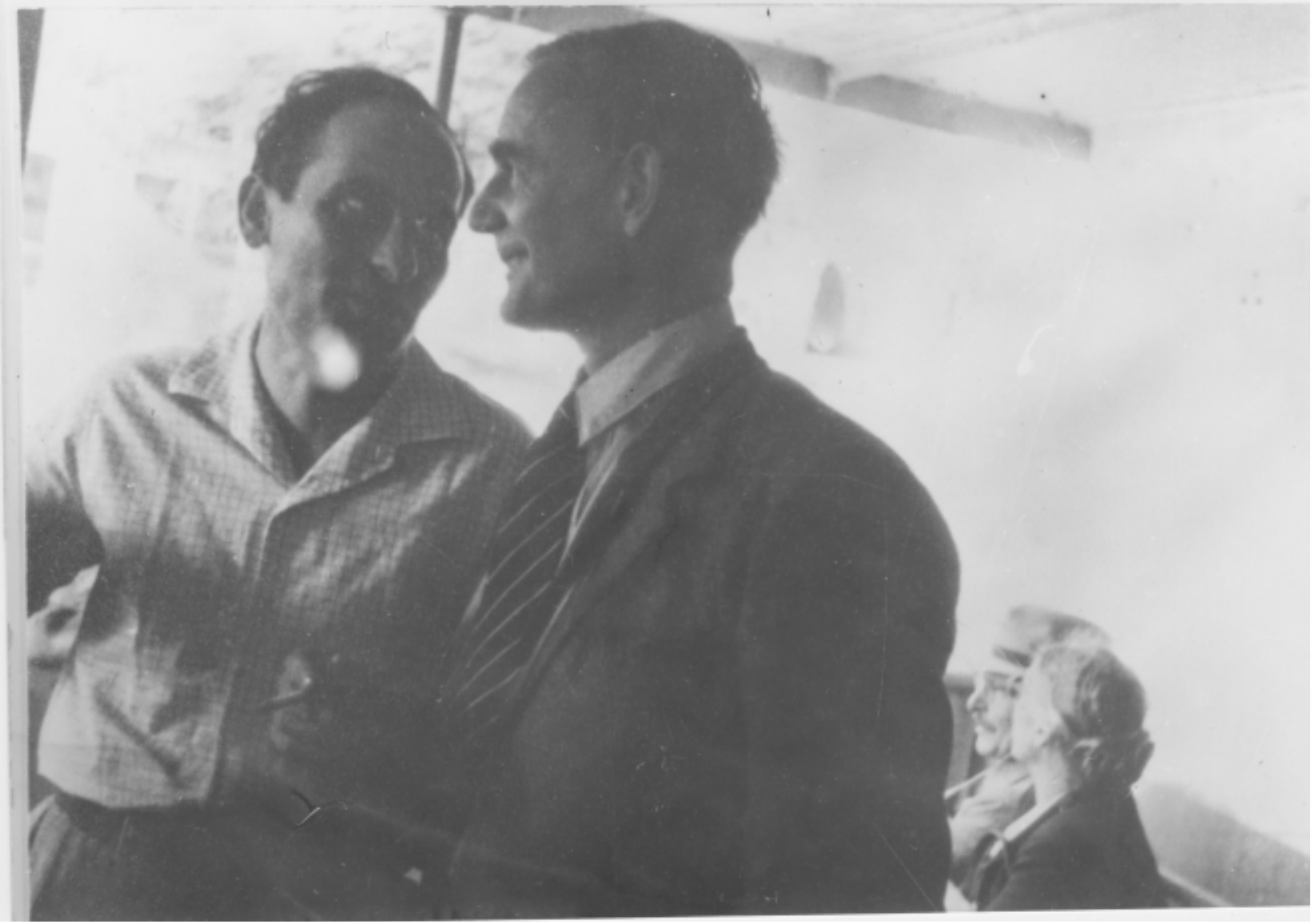}
\caption{Como Conference, 1949. Occhialini (left) with Powell (Fondo Della Corte, Science Library, Scientific and Technological Hub, University of Florence).}\label{fig7}
\end{figure}

Unfortunately, at the end of the 1930s, this golden age of Florentine physics came to an end: in 1937, Bernardini left Florence for the chair of the University of Camerino. Occhialini, after his return in 1934 in Florence, in 1937 left for Brazil to escape Fascist Italy  sizing the opportunity of Gleb Wataghin invitation. In Brazil, Occhialini contributed to the birth of the Brazilian school of physics in S\"ao Paulo. He returned to Europe in 1944 and collaborated with the Brazilian Cesare Lattes and with Cecil Powell on the discovery of the pion in 1947, Fig.~\ref{fig7}.
Daria Bocciarelli moved to the Istituto Superiore di Sanit\`a in Rome in 1938, where she played a very important role. Racah moved to Pisa in 1937, called to Pisa University after winning the second Italian selection for professorships in Theoretical Physics. After a period in the United States at Stanford, Emo Capodilista abandoned his scientific career.
Added to all this in 1933 was the death of Antonio Garbasso, at the age of 62, and his replacement by Laureto Tieri, who was Director from 1933 to 1948. The atmosphere at the Institute had changed, {\it the spirit of Arcetri} was no longer there. Perhaps these young people would have gone anyway, but certainly the change of the human environment at the Institute was largely responsible for it. Concerning Tieri, Della Corte's opinion is worth quoting: 
 {\it He was an old-school professor, who delivered his lectures with the dignity and nobility typical of the late 19th century, but unfortunately the course content was also from the same era!}
 
 By the end of the 1930s, Manlio Mand\`o, Michele Della Corte and Giuliano Toraldo di Francia had graduated from Arcetri, and, together with Simone Franchetti and Nello Carrara, they were to contribute to the rebirth of Florentine physics in the post-war period (see Table~1). 


\begin{table}[h]
\begin{center}
\caption{Thesis title and year of graduation of  Simone  Franchetti,  Manlio  Mand\`o,  Michele Della Corte and Giuliano Toraldo di Francia}
\end{center}
\label{tab1}%
\begin{tabular}{@{}lll@{}}
 & Thesis year &Thesis title  \\
 &&\\
S. Franchetti   & 1933  & Thermal frequencies and interatomic forces \\
 M. Mand\`o   & 1934/35 & The valve amplifier method
 \\
    &  & 
for the study of elementary particles
 \\
 M. Della Corte   & 1938  & On the thermionic effect  \\
 G. Toraldo  di Francia& 1939  & On the motions of a viscous liquid \\
  & &  symmetrical with respect to an axis \\
\end{tabular}
\end{table}

The role of Arcetri and its researchers during that period was recognised by the European Physical Society, which declared Arcetri a Historical Site for Physics in 2012. The research carried out during those years had very important consequences for the development and applications of Fermi-Dirac statistics and in experimental research in relation to 
cosmic rays. Remember that cosmic rays were fundamental for the development of detection techniques that were later used and further refined in modern experiments in elementary particle physics. There was a second particularly fertile period in the history of Florentine physics, as we shall see later, and this was in the 1960s.

\section{During the war}

\noindent

The life of the physics community of Arcetri during the years of the war can be better understood through some relevant episodes.

In academic year 1937/38, Franchetti became Assistant Professor and lecturer in Theoretical Physics in Florence, where he began his studies on the physics of the nucleus, focusing particularly on the interaction of $\gamma$-rays with matter, and later on the spectra of $\mu$ leptons or, as they were then called, mesons.

In 1937, Tito Franzini had arrived from the University of Pavia, and Vincenzo Ricca, from the University of Messina. Tito Franzini moved as Assistant Professor to Florence to take over the position vacated by Mand\`o who, in the meantime, had become Assistant Professor in Palermo, where Emilio Segr\`e had been called as visiting Professor from 1935.

After the departures described at the end of the previous chapter, the {\it highly lerned physicist} \citep{dc99} Simone Franchetti, the last to remain at the Institute from the time of Bernardini, was also expelled from the Florentine university due to the racial laws, because, although his mother was Catholic, his father was Jewish. A few months earlier, on 1 November 1938, he had been appointed Assistant Professor. The Rector, Arrigo Serpieri, wrote to Franchetti on 24 January 1939, informing him that he was to be released from service on 14 December 1938 because, according to the Ministry,  {\it A person born of mixed marriages who does not profess any religion, even if he   never made any manifestation of Judaism, cannot be considered Aryan, as the law requires positive membership of a religion, which must, however, not be Jewish} \footnote{Letter of the Rector A. Serpieri to S. Franchetti, 24/1/1939 (Fondo Della Corte, Science Library, Scientific and Technological Hub, University of Florence).}. And Franchetti had declared himself non-confessional and, therefore, lacked {\it positive affiliation to a religion other than Judaism}. Around forty lecturers of Jewish origin were expelled from the Florentine university in 1938\footnote{see Introduction by P. Guarnieri in \citep{guarnieri}  and pertinent quotes.}. Initially, the Director Laureto Tieri had allowed him to continue his research work at the Institute in the evenings, from 9 p.m. to past midnight. {\it Then, perhaps due to a informer or the presence of fascist elements in the Institute, this authorisation was revoked} \citep{dc99}. 

In addition to the Director Laureto Tieri, a number of young graduates remained at the Institute of Physics, including Carlo Ballario and Della Corte, who, together with Tito Prosperi in 1940, carried out the experiment on the absorption of cosmic rays under rock in the Tuscan-Emilian Apennines near Castiglione dei Pepoli. The experiment took place in a shaft of the tunnel of the Florence-Bologna railway, which descended down a slope to a depth of 200 metres, to the underground station called Precedenze.  The aim of the experiment was to detect the particles that make up cosmic rays, after passing through the layers of matter of the mountain, using Geiger  M\"uller  counters and an electrical coincidence circuit.  The apparatus was mounted on the carriage of the well, and this allowed measurements to be taken at various depths  \citep{bdcp1939}.

Della Corte then left for military service, first in the infantry corps and then, from 1942, at the Cascine Air Warfare School, where he met air force captain Italo Piccagli, a staunch anti-fascist. In the period immediately preceding 8 September 1943, Piccagli offered Della Corte the opportunity to transfer the instruments and apparatuses of the Meteorology and Air Navigation Laboratories to Arcetri to save them from requisition by the German army. The equipment was hidden in the Institute of Physics. A few months later, Della Corte and Ballario, again at the suggestion of Captain Piccagli, joined Radio CORA, a clandestine radio station promoted by the Radio Commission of the Action Party. The radio, which had the task of transmitting information to allied commands and partisan troops, had several bases in the city, including the Institute of Physics in Arcetri.

The Institute was then searched by the German SS, who had been alerted to the presence of air force material on the premises. Fortunately, the material had been well hidden and the Germans only took away a few items. Della Corte, who was forced to accompany the SS officer during the search, managed to find out who was behind the information given to the German army. It was a fellow physicist from the University of Florence who collaborated with the German army, Ivo Ranzi. The Florentine episode, in which Ivo Ranzi demonstrated his collaboration with the German Command, was not the only one.  In Rome, too, Ranzi was suspected of having passed on to the Germans information he had received from the Director of the Physics Laboratory of the Istituto Superiore di Sanit\`a, Giulio Cesare Trabacchi, concerning the availability of radium, at the Institute \citep{sanita}. 

For all these reasons, after the end of the war, Ranzi was subjected to purging proceedings, which ended in the first instance with a proposal of dispensation from service, cancelled on appeal by the Council of State following the application of Decree Law no. 48 of 7 February 1948, which had declared the proceedings pending to be extinct. Ranzi at that point
should have resumed service at the Florentine university but, on 22 March 1948, the Faculty of Mathematical, Physical and Natural Sciences of the University of Florence ruled against his reinstatement:
{\it considering that, after 8 September 1943, he went into the service of the German army, when his duty was to resume his position as professor, as the Faculty wished, and that he also took action that damaged the Institutes of Arcetri,} (Author's note: the Faculty) {\it   is unanimous in its agreement to deny the approval of professor Ranzi's return to occupy the chair of Advanced  Physics at the University of Arcetri}  \footnote{Resolution of the Faculty of Science, ASUF, Personnel Section, personal files of lecturers, Ranzi Ivo, file 3043 series A}. As a consequence, the Rector denied Ranzi's return to the chair of Advanced Physics and issued a negative opinion on Ranzi's request for leave of absence, referring him to the Disciplinary Board for breach of official duties and for 
having performed acts detrimental to the dignity of his role as professor. The situation was resolved by the Ministry of Foreign Affairs and the Ministry of Education, which placed Ranzi at the disposal of the Ministry of Foreign Affairs until the end of 1953, for special research assignments in Argentina. He resumed his service as a professor in Florence in March 1954.  On 1 January 1959, Ranzi left the university and moved in Rome to the Scuola Superiore di Telegrafia e Telefonia, holding the chair of radiotelegraphy and radiotelephony until his retirement on 1 November 1978.

The experience of Radio CORA ended tragically on 7 June 1943; the Nazis, having located the radio in the premises in Piazza d'Azeglio in Florence, stormed in. Luigi Morandi, who was on duty at the radio transmitter in the attic, managed to fight back and kill a German, but was wounded and died a few days later. Despite being tortured, Piccagli and the lawyer Enrico Bocci\footnote{The man, together with Piccagli, behind Radio CORA.}, took the entire responsibility for the organisation, exonerating the others. Piccagli was killed by German soldiers in the woods of Cercina (in the hills above Florence) on 12 June 1944, together with other anti-fascists. Enrico Bocci was probably killed by the SS, but his body was never found.

	Della Corte and Ballario were saved by the sacrifice of their comrades and because they were not on radio duty that day. From 1944 to 1947, Ballario was Assistant Professor at the University of Bologna before moving to Rome, where he collaborated with Amaldi and Bernardini's group. Della Corte continued his academic career in Florence.
	
Simone Franchetti's expulsion from the university, due to his Jewish origins, ended on 28 December 1944 when he resumed service at Arcetri.

Mand\`o was called to arms and sent to Libya in 1939.  In December 1940 Mand\`o was badly wounded in battle and captured by the British army. He was operated on and treated by the British in the military hospital in Alexandria and then transferred to prison camps in India. He finally returned to Italy on 1 July 1946. During his years of imprisonment, taking advantage of the fact that the Italians were able to benefit from a progressive easing of the rigours of imprisonment up to a regime of semi-liberty after the armistice of September 1943, Mand\`o worked with whatever means were available, along with others, to organise university courses for Italian prisoners who had abandoned their studies when called up for war. It was in this context that some of Mand\`o's (later improved) didactic texts published in the years following his return to Italy from captivity were born. These would become a point of reference for the study of General Physics I for entire generations of Physics students and others in various Italian universities \citep{mandFG1es,mandFG1}. In September 1946, upon his return to Italy, he resumed his service at the University of Bologna, where he had been Assistant Professor before being called up for military service.

The Institute of Physics was also used by the Allied  during the period of occupation: university-level courses for the military troops stationing in Florence where organised there by the American Command. For that purpose, the Americans had also brought electronic material to the Institute for educational purposes. In addition, some men of the American counterintelligence were installed  in one of the two terraces of the Institute  to intercept radio broadcasts of fascists who fled to the North \citep{dc99}.

\section{After the war until the 1960s}

In this section we retrace the birth of the three main areas of research in Florence in the 1960s: high energy physics, which saw the transition from experiments with cosmic rays to those conducted with particle beam accelerators, physics of the nucleus, where, again, there was a transition from the use of radioactive substances in research to the use of accelerator beams to bombard nuclei for activation, and finally theoretical physics. In this process Gilberto Bernardini plaid a worthfull role thanks to his expertise both in cosmic rays physics, particle physics and later accelerators.

\begin{figure}[h]
\centering
\includegraphics[width=0.7\textwidth]{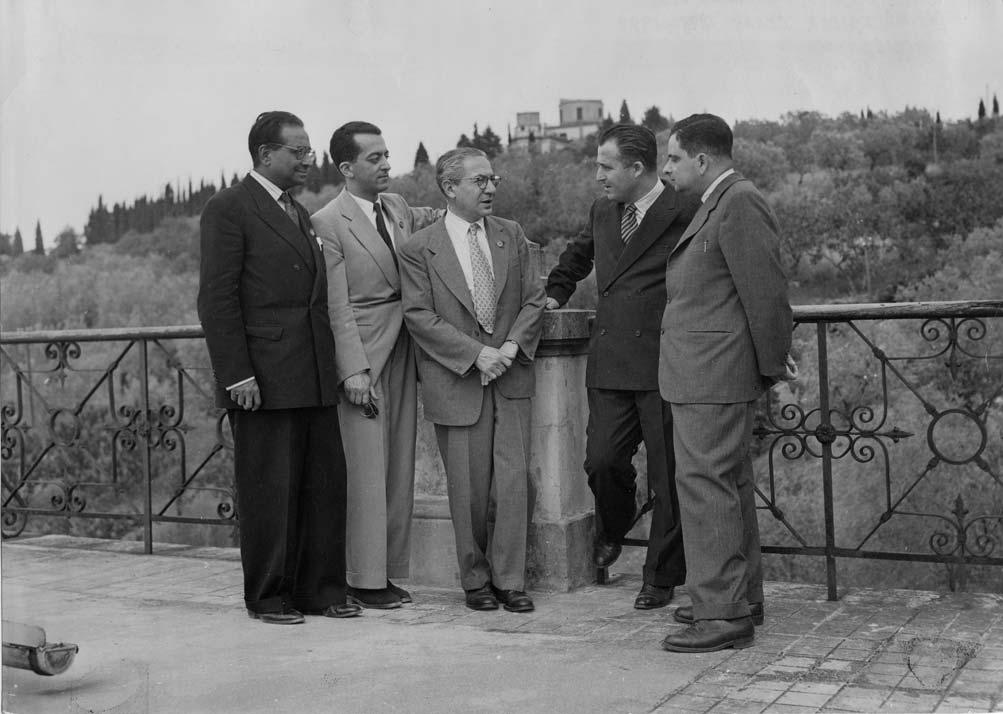}
\caption{From the right Franchetti, Mand\`o, Isidor Rabi and two other visitors on the Arcetri terrace. Arcetri Observatory can be seen in the background before the dome of the Amici Telescope was installed (Department of Physics and Astronomy, Scientific and Technological Hub, University of Florence).}\label{fig9}
\end{figure}

In the years of reconstruction and reorganisation of research after the Second World War, in June 1950, the Fifth General  UNESCO conference was held in Florence.  Despite the fact that nothing doing with European collaboration on nuclear research had been put on the agenda, a general resolution of that sort was voted by the  participants following a suggestion of Isidor Rabi who was a member of the US delegation \citep{pestre1984}.  CERN would come into being in 1954. In Fig.~\ref{fig9} we can see Rabi visiting the Physics Institute in Arcetri.

In 1951, on the initiative of Edoardo Amaldi and Gilberto Bernardini and with the support of Gustavo Colonnetti, President of the CNR, a new research agency, the Istituto Nazionale di Fisica Nucleare (National Institute for Nuclear Physics), was born with the aim of coordinating theoretical and experimental research activities in nuclear physics and cosmic rays. Its first four Sections were those of Rome, Padua, Milan and Turin. A Subsection was created, in Florence, in 1952, and was directed by Franchetti until 1966, by Renato Angelo Ricci in 1967/68 and then by Mand\`o until 1972, when it was transformed into a Section. The first Director of the Section was Giuliano Di Caporiacco. As for the Physics Institute, Tieri remained its Director until 1949, when he was succeeded by Simone Franchetti, who directed it until 1977.

\subsection{Experimental physics}

\begin{figure}[h]
\centering
\includegraphics[width=0.7\textwidth]{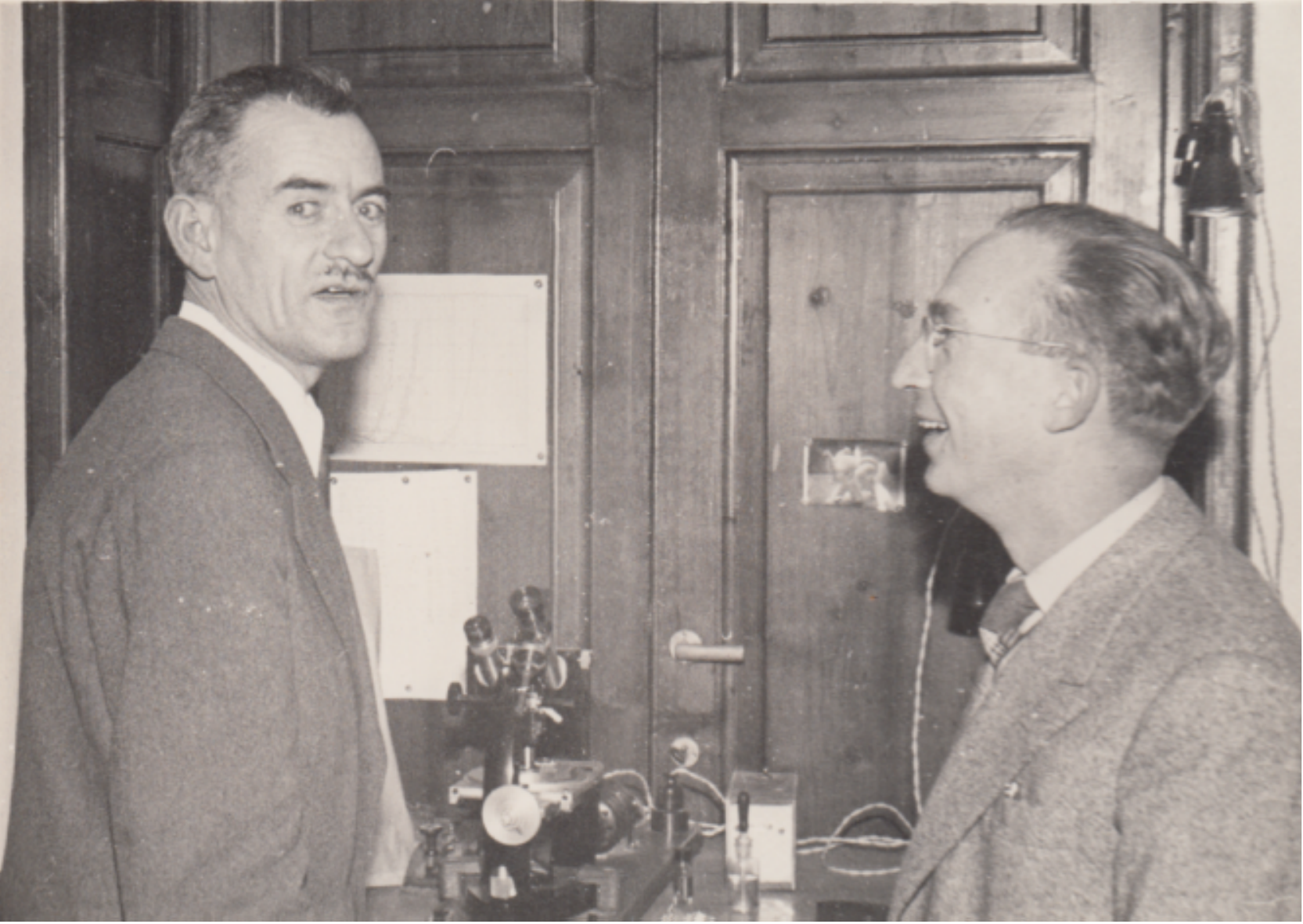}
\caption{Della Corte (right) with Louis Leprince-Ringuet in Florence in 1954 (Fondo Della Corte, Science Library, University of Florence).}\label{fig8}
\end{figure}

The Arcetri school's research in cosmic rays was continued by Della Corte (Fig.~\ref{fig8}), who, in 1950, using a CNR grant followed by another from the Della Riccia Foundation, went to Paris to work with Louis Leprince-Ringuet's group at the \'Ecole Poly-technique, specialized in nuclear plates to learn this  technique. Upon his return, he created the Florentine {\it plate group} \citep{dcl,Cartacci2014}, a team performing particle physics experiments with nuclear emulsions. Della Corte studied the phenomenon of trace formation and developed a model of it \citep{DellaCorte1952PhotometricMO,Corte1953TheGD,Corte1954,dc1956,Bizzeti1958PhotometricAO,Bizzeti1959OnTT}. The group's first research was in cosmic rays to determine the charge and mass of the particles. Up to the mid-1950s, Marco Giovannozzi \citep{Corte1951AnIA,Caporiacco1956CloudCS}, Mand\`o \citep{Mand1952OnTE,ms1953}, Franchetti and Tito Fazzini \citep{Corte1955AnelasticSO} also collaborated on cosmic rays. Fazzini had graduated with Franchetti in 1947 on the energy spectrum of $\mu$ \citep{Corte1947TheMS}.


The Florence group, which was joined over the years by Anna Maria Cartacci and Pier Giorgio Bizzeti, among others, together with a number of observers who were particularly well trained in scanning events under the microscope, after this first period of internal reconstruction, switched from research on cosmic rays to research with accelerators.  In particular  it later steered its research by collaborating with groups from other Italian universities at  CERN, the new European facility for particle and nuclear physics.

After the study of positive pion-proton scattering at 8.3 MeV to determine phase shifts \citep{Ferretti1957ProtonSA}, the Florence-Genoa-Turin collaboration studied 25 GeV proton interactions at CERN's Proton Synchrotron, determining the characteristics of these interactions \citep{Bizzeti1963OnTI,MarzariChiesa1963OnTI};  the Parma-Florence collaboration on the other hand studied the final products in the interactions of strange mesons in nuclear emulsions \citep{Quareni1965TheRO}. In 1964, the group abandoned the technique of nuclear emulsions and switched to the analysis of bubble chamber frames in experiments at CERN. The transition was not hard: observers of microscope plates were easily and profitably transformed into observers of bubble chamber images. In 1965, Pier Giorgio Bizzeti left the group to concentrate on nuclear physics.

At the end of the 1960s, Della Corte, having become increasingly interested in the application of physics to medicine, switched to nuclear medicine, abandoning elementary particle research, which required large-scale collaborations in which the role of the individual researcher became increasingly anonymous. However, Giuliano Di Caporiacco and, a little later, Giuliano Parrini joined the group.  Techniques were refined and a number of projectors that displayed the frames taken at CERN in the bubble chamber on the measurement tables were purchased. Digitised readers providing the coordinates of the traces were built on site. In addition, large event reconstruction programmes began to be used, exploiting the computers at the CINECA computer centre in Casalecchio di Reno (BO). Big international experiments were 
launched at that time. In collaboration with the groups in Bologna, Bari and the Institut de Physique Nucl\'eaire in Orsay, the Florentine group published results on the research into new particles in resonant pion systems, produced in interactions of the 5.1 GeV positive pion beam of CERN's Proton Synchrotron in a bubble chamber filled with deuterium \citep{Armenise1967Ao2PI,Armenise1968RI,Armenise1969A3PI,Armenise1970QuasitwobodyPI}. The research was then extended by the three Italian groups with the 9 GeV beam. A new collaboration with the groups of Milan, Bologna and Oxford continued studying the production of
new states in interactions of $\pi^-$ of 11.2 GeV in a bubble chamber filled with hydrogen. The group would later return to the emulsion technique in the search for particles containing a $c$-type quark (charm), carried out at CERN in the 1970s.

At the end of the 1950s, Manlio Mand\`o, after moving from Bologna to Florence as an Assistant Professor of Experimental Physics and having carried out research on cosmic rays in the immediate post-war period, succeeded in setting up an experimental group to work on nuclear physics. In the years that followed, this group became of international importance thanks partly to its fruitful contacts with centres of excellence for nuclear physics in the United States, Japan and especially Germany. In addition to Manlio Mand\`o, Tito Fazzini, Piergiorgio Bizzeti (after the initial period with the plate group), Anna Maria Bizzeti Sona, Mario Bocciolini, Giuliano Di Caporiacco (who later joined the plate group) and, from 1962-63, the new graduates Pietro Sona, Paolo Maurenzig, Nello Taccetti and Paolo Blasi were part of the nuclear physics group of the INFN subsection in Florence. In the early 1960s, this group turned to experimental nuclear physics with accelerators, using a  PN400 Van de Graaff accelerator, which provided a terminal voltage of 400 kV and could accelerate protons and deuterons to be sent to a target, with the aim of producing neutrons and $\gamma$-rays. The accelerator was housed in a purpose-built bunker next to the Institute of Physics \citep{Taccetti2017}, and the fine-tuning of the accelerator was taken care of by the group of technicians and researchers.

\begin{figure}[h]
\centering
\includegraphics[width=0.7\textwidth]{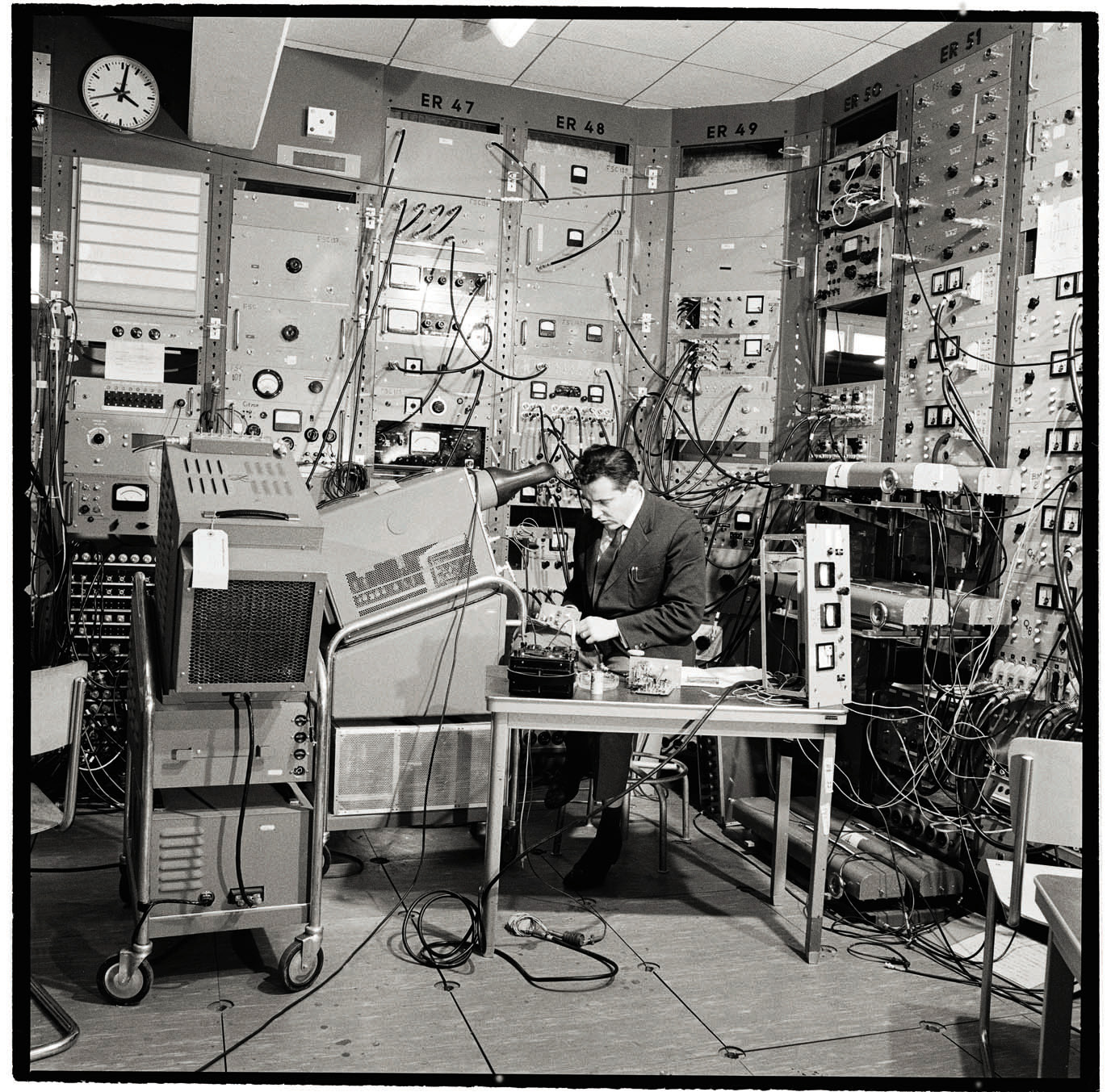}
\caption{Tito Fazzini in the control room of the experiment at CERN (CERN Archive).}\label{fig9b}
\end{figure}

The main activities carried out were the production of isomeric states \citep{Ademollo1960OnTI} and the measurement of photo-production cross-section fluctuations \citep{Mand1962The7LipA,Bizzeti1963OnTI,Bizzeti1965ErsicsonsFI}.  The group also worked on the development of new detectors.  When Ricci moved to the University of Florence in 1965, called by Franchetti and Mand\`o, he found the aforementioned group of young experimental physicists, which was studying the use of new Silicon-based detectors to detect electrons and $\alpha$ particles, and new Germanium-based detectors to detect $\gamma$-rays. It was also busy designing a vacuum chamber for the target and the detectors \citep{bbms1966,Benvenuti1965AVV,Blasi1966ASF}. Once the technical difficulties involved in their proper use had been overcome, these detectors made it possible to achieve much better energy resolutions than could be obtained with thallium-doped sodium iodide detectors. The radioactive sources to be studied were obtained using a 14 MeV neutron beam produced by the Van de Graaf accelerator at 400 kV. At the end of 1966, Ricci proposed that the group participate in nuclear spectroscopy research at the Legnaro Laboratory (Padua), which was to become the INFN's Legnaro National Laboratories dedicated to nuclear physics in 1968. This marked the start of the Florentine nuclear group's long collaboration with the Legnaro Laboratory, where a 5 MV Van de Graaf was in operation and spectroscopy measurements were carried out [see review article \citep{Ricci1969The1P}]. The Florentine group also established collaborations with other important European nuclear physics laboratories, such as those in Saclay, Heidelberg and Munich. 
From 1957 to 1961, Tito Fazzini was employed at CERN (Fig.~\ref{fig9b}), where he took part in important experiments. Particularly noteworthy is that carried out in August 1958, when, together with Giuseppe Fidecaro, Alec Merrison, Helmut 
Paul and Alvin Tollestrup, he proved that one pion in 10,000 decays into an electron-neutrino in accordance with the theory of weak interactions, originally introduced by Fermi in 1933
\footnote{For an overview of this group of experiments see \citep{Fidecaro:2017aee} }.  Fazzini returned to Florence in 1962, where he focused on research into nuclear physics with Mand\`o's group.

The PN400 accelerator was decommissioned at the end of the 1960s. The Florentine group managed to obtain the KS3000 electron accelerator, a 3 MV Van de Graaf, from INFN. The KS3000 had been the injector of the Frascati Electron Synchrotron, the activity of which had begun in the mid-1950s and ended in 1968-69. The accelerator arrived in Florence in 1971 and was housed in the bunker of the decommissioned PN400. The electron accelerator was transformed into a positive ion accelerator, with the important contribution of Tito Fazzini, Giacomo Poggi and Nello Taccetti and the group's technicians. The new accelerator, renamed KN3000, was used in the 1970s to perform nuclear spectroscopy measurements, two parity violation experiments in nuclei. Its career came to a close with participation, under the leadership of Pier Andrea Mand\`o, in the initial phase of the programme of nuclear physics applied to the environment and cultural heritage \citep{Taccetti2017,mand2014}. This initial activity gave birth to the group that now works at the Laboratory of Nuclear
Techniques Applied to Cultural Heritage (LABEC) at the Scientific Hub of the University of Florence in Sesto Fiorentino.

Lastly, we would like to mention the physics activities that were born and developed in Florence around the figures of Nello Carrara and Giuliano Toraldo di Francia, in the laboratories of the CNR Microwave Centre and of what was to become the Institute of Advanced Physics of the University of Florence.

After his long experience as a teacher and researcher at the Livorno Naval Academy, Nello Carrara founded the CNR Microwave Centre in 1946 with the aim of studying the applications of physics to radio propagation; it was in this laboratory that the first Italian radar, intended for use in the Italian Navy, was designed and built. In 1954, he became Professor of the course in Theory and Technique of Electromagnetic Waves at the Naval University Institute in Naples, before moving to the University of Florence in 1956. In the 1950s Carrara contributed his experience as a founding father of Italian electronics to the Frascati synchrotron by collaborating with Federico Quercia and Mario Puglisi on the construction of radio-frequency cavities to accelerate electrons. In the years that followed, the Microwave Centre expanded to cover various experimental activities in the study of matter and applications of lasers and, in 1968, it became the Electromagnetic Waves Research Institute (Istituto di Ricerca delle Onde Elettromagnetiche - IROE) \citep{cdm2021}.

Giuliano Toraldo di Francia, an important figure in physics and of Florentine culture in general, was first Lecturer from 1951 to 1958 and then Professor of Optics at the University of Florence. He founded the Institute of Radiation Physics, which later became the Institute of Advanced Physics. After extensive research into classical optics and evanescent electromagnetic waves, he began a new research activity, immediately realising the importance of the discovery of the laser. He directed his group's research in this sector, with the development of the first fibre optics in Italy and of laser spectroscopy, initiating the birth of research into the structure of matter in Florence. Toraldo also made a significant contribution to studies in the philosophy of science \citep{cdm2021}.

\subsection{Theoretical Physics}

To conclude, we are going to look at the last area of research present in Florence at that time, that of theoretical physics.

After his initial studies of nuclear physics and cosmic rays, from the 1950s onwards, Franchetti concentrated on the theoretical study of condensed states, especially liquids. He was among the pioneers who tackled problems of considerable interest such as that of liquid helium \citep{Franchetti1954RleOE,Franchetti1955ProblemsOF,Franchetti1955TentativeOT,Franchetti1960SomeRO,Franchetti1961OnTS,Franchetti1964OnTF}.

After Persico's departure in 1930, the chair of Theoretical Physics had remained vacant until 1958, when Giacomo Morpurgo was appointed. Morpurgo held it until 1962. The Theoretical Physics course had been held from 1944 until that year by Franchetti. With Morpurgo's arrival, the Institute of Theoretical Physics was born\footnote{Despite the brief period that Morpurgo spent in Florence, Claudio Chiuderi \citep{Chiuderi1961CoulombPO}, Emilio Borchi, Giovanni Martucci and Mario Poli graduated with him. Emilio Borchi, together with Silvio De Gennaro, formed the first nucleus of the structure of matter theoretical group at the end of the 1960s. Claudio Chiuderi, on the other hand, focused on plasma physics some years later. }.

Morpurgo moved to Genoa in 1963 and Raoul Gatto, who also became Director of the Institute, was assigned the chair of Professor of Theoretical Physics\footnote{On Raoul Gatto see  \citep{Casalbuoni_Dominici_2018,battimelli}}. 

Raoul Gatto, born in Catania on 8 December 1930, was accepted at the Scuola Normale Superiore in Pisa in 1946. In 1951, he graduated with first-class honours with a thesis on shell models of nuclei, prepared under the guidance of Bruno Ferretti, who held the chair of Theoretical Physics in Rome at the time, and Marcello Conversi, who taught experimental physics in Pisa and was his supervisor. In the same year he obtained a diploma from the Scuola Normale with honours with a thesis on statistical theories of nuclei, supervised by Prof. Derenzini. After his thesis, Gatto moved to Rome and became assistant to Bruno Ferretti,  studying weak hadron decays and associated angular distributions. He found a very open international environment in Rome and considerable scientific stimuli, thanks to Edoardo Amaldi and the legacy of Enrico Fermi's school. In 1956 Gatto obtained the "Libera docenza" in Theoretical Physics and moved to the Radiation Laboratory in Berkeley, California, where he remained until 1957. At Berkeley, where Louis Alvarez's group was discovering numerous new particles using a bubble chamber, Gatto continued his work on the phenomenology of hyperons, the symmetries of weak interactions and the consequences of the violation of parity in weak interactions, which had just been discovered by Chien-Shiung Wu and her collaborators. Gatto suggested the conservation of CP symmetry (a product of charge conjugation and parity), independently of Lev Landau and T.D. Lee and C.N. Yang.  Fitch and Cronin's experiment in 1964 showed that CP symmetry is also violated.

In the early 1960s, Gatto became a Full Professor in Cagliari, but also spent part of his time at the Frascati Laboratories. As one of the leading theoretical experts in Quantum Electrodynamics, he participated in the research for the first particle collider, or Storage Ring (AdA), the well-known electron-positron machine devised by Touschek, and in the research for the subsequent ADONE. Famous from that period is the work by Gatto, in collaboration with Nicola Cabibbo, on the physics of colliding beams, known among researchers as the "Bible"\citep{Cabibbo:1961sz,bibbia}.

\begin{figure}[h]
\centering
\includegraphics[width=0.5\textwidth]{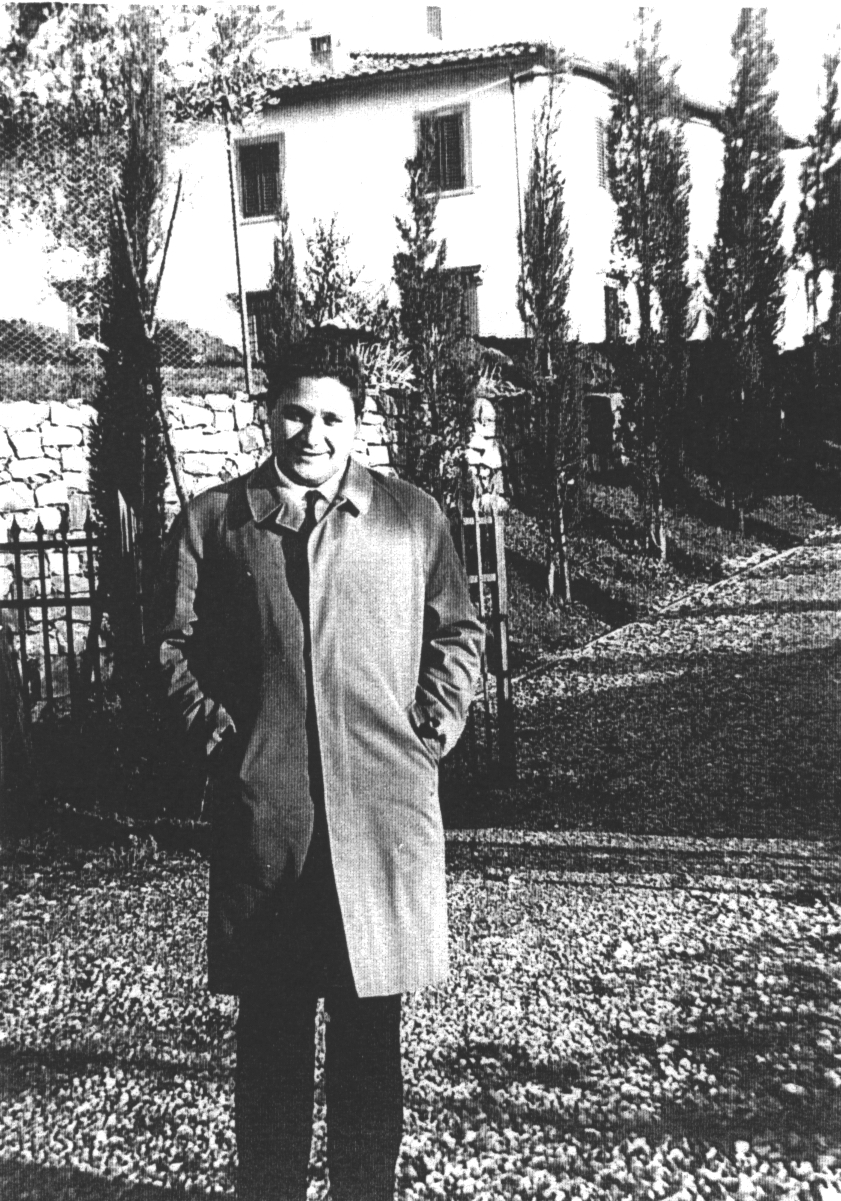}
\caption{Raoul Gatto in Florence in the 1960s (courtesy of the family).}\label{fig10}
\end{figure}

In Florence, Gatto (see Fig.~\ref{fig10}) decided to create a group of young theorists, which he organised in the style of the famous Landau school.  There he found the young Marco Ademollo, Giorgio Longhi and Claudio Chiuderi and brought to Florence some students who had done their theses with him at the university in Rome.

Guido Altarelli and Franco Buccella, who had done an important thesis with Gatto entitled {\it Calculation of cross sections for the emission of a photon in the 
 $e^+e^- \to e^+e^-\gamma$   collision} \citep{Altarelli1964SinglePE}, came to Arcetri, along with                                                                                                      Giuliano Preparata who had done a thesis with him on the determination of the spin of bosons. Another young man from Rome who moved to Florence was Luciano Maiani, who graduated from Rome in 1964 with an experimental thesis on semiconductor detectors, supervised by Mario Ageno of the Istituto Superiore di Sanit\`a.  After graduation, Ageno agreed to give him a scholarship from the Istituto Superiore della Sanit\`a and allowed him to work not only in Rome but also in Florence with Gatto. The only Florentine student who graduated with Gatto in this early period was Gabriele Veneziano. Veneziano's time in Florence after graduation was brief but intense. It was in Florence that Veneziano began his collaboration with Ademollo on sum rules, which led him to the formulation of the internationally acclaimed model that bears his name in 1968  \citep{Veneziano:1968yb}.

Gatto, who had learned the importance of symmetries applied to particle physics at Berkeley, recalls the research activity of his Florentine period as follows:  {\it Marco Ademollo was one of the most continuous collaborators within the group. It is interesting that one of Marco's initial interest lay in a problem with rather arduous formal aspects, but which remained somewhat marginal: the problem of determining spin} \citep{Ademollo1963MethodsFD,Ademollo1964CompleteST}. {\it In those years, the discovery of new particles and resonances followed one another incessantly, and rapid and complete methods were needed to determine their spins from various angular correlations. This line of research was later taken up by Giuliano Preparata }\citep{Ademollo1964TESTSFS}{\it .  Marco's interest later shifted, as did that of the whole group, to the properties of symmetry. The non-renormalisation theorem was deduced by means of an algebraic method }  \citep{Ademollo1964NonrenormalizationTF} (Known as the Ademollo-Gatto theorem, Author's note). 
{\it The collinear subgroups were one of the later interests. Franco Buccella was initially interested in this line} \citep{Buccella1965TheCG}. {\it  Later, in addition to Ademollo, Longhi and Veneziano also took an interest }\citep{Ademollo1967MixingSF}.

{\it  Another interest at the time was in sidewise dispersion relations, on which, in addition to Ademollo} \citep{Ademollo:1969wn}, {\it Casalbuoni and Longhi also worked later on} \citep{Casalbuoni:1969gv}. 
{\it  Longhi's activity was long and constant.  In addition to that already mentioned, he worked extensively on electron-positron radiative corrections. This subject had already been addressed by Altarelli and Buccella in their dissertation in Rome. Longhi worked on it with Altarelli, Celeghini, and De Gennaro} \citep{Altarelli1967TheoreticalCF}. {\it Pino Furlan  collaborated on this research} \citep{fgl1964}.

{\it Again in relation to electron-positron, Longhi worked on baryon resonances, together with Casalbuoni.  The subject had already been dealt with by Celeghini, after his work on $\eta_0$} \citep{Celeghini1964PossibleDO,Celeghini1964PossibleMT}. {\it  Celeghini, was one of the first to come to Florence from outside and later participated in the work on current algebra saturation} \citep{Celeghini1966LocalCR}. {\it Borchi's activity deserves special mention, especially with regard to his work on what was the first classification of mesonic resonances obtained by conjecturing that these resonances were part of  a $p$-wave multiple} \citep{Borchi1965AssignmentOH}. {\it The essentials of this classification, which was very conjectural at the time, turned out to be correct over time and have been accepted, as in the recent works of 't Hooft, Maiani, et. al. Maiani, Preparata, Longhi and Veneziano expanded these studies shortly afterwards. At that time, there was great uncertainty surrounding mesonic resonances. It was a matter of rummaging through the Particle Data and seeing if an attempt at classification could be made. The introduction of $p$-waves seemed necessary. It should be mentioned that Borchi had also contributed to work on muons (radiative capture) by studying the possible use of data to verify the properties of weak currents}  \citep{Borchi1965RadiativeMC}.  {\it The group worked extensively on the search for symmetry properties of strong interactions, in an attempt to exploit the phenomenological success of SU(6), which later proved difficult to formulate theoretically. Leading the way in this theoretically complex effort were Altarelli, Buccella, Maiani and Preparata} \citep{Altarelli1965TheVA}. {\it  
Buccella and, separately, Veneziano took an interest (together with Okubo) in the tricky problem of Schwinger's terms (then almost unknown) } \citep{Buccella1966NecessityOA}.
{\it 
Mention should also be made of Altarelli and Buccella's interest in the non-leptonic decays of hyperons, which led at the time to the suggestion of relations between amplitudes, later found to be correct, despite the uncertain theoretical foundation SU(6) }\citep{Altarelli1965SU6A}. {\it 
Gallavotti, who possessed a vast mathematical knowledge, wanted to tackle problems in statistical mechanics and clarify delicate issues, such as the structure of gauge invariance in electrodynamics. Alongside him, others with a clear mathematical leaning should also be mentioned, such as Giusti, Mosco, and especially Da Prato. The mathematical interest of these young people came from theoretical physics and their participation in discussions and seminars was constant. They later had brilliant careers in mathematics} \footnote{Gatto on theoretical physicists of the Florentine period, private communication to Casalbuoni.}. Maiani \citep{maiani} and Preparata \citep{preparata} also speak about this Florentine period in their books. 

In the 1968 Europhysics conference, Harry Lipkin coined the name New Florentine Renaissance for this Florentine school created by Gatto.

In 1966, Gatto, who taught Theoretical Physics, decided to expand the group and began a recruitment campaign among third-year students, attending the examinations of Institutions of Theoretical Physics, a course taught by Ademollo, and proposing a thesis on Theoretical Physics to those students who seemed most promising. A number of young people were recruited in this way, including one of the authors of this work (RC).\
Marcello Colocci arrived in Florence in 1966. After a period of research at CERN, from 1968 to 1970, he would return to Florence, where, together with Ruggero Querzoli, he would create a group that would deal with the structure of matter, training many young researchers over the years that followed.

On the whole, Gatto's students were truly exceptional, as is shown by the success of their scientific careers, but certainly the atmosphere and his teaching contributed in no small measure to what has been considered the only true school of theoretical physics ever to have existed in Italy. There were many scientifically important results, probably the best known of which is the aforementioned Ademollo-Gatto theorem on certain properties of weak interactions (non-renormalization for the strangeness-violating vector currents to first order in the symmetry breaking  interactions).

In academic year 1967-68, Gatto went on leave to Geneva. In academic year 1968-69 he moved to Padua and again to Rome. In the 1970s, he was called to the University of Geneva to occupy the prestigious chair previously held by Ernst St\"uckelberg. He taught there until 1995.

At the end of the 1960s, it was practically impossible to keep the young Roman researchers, all of whom held precarious positions such as scholarships and other temporary contracts, in Florence. So, more or less at the same time, all the Romans, Altarelli, Maiani and Preparata in 1968, Gallavotti in 1969 and Buccella in 1970-71, went to the United States or CERN for one or two years for research, and then returned to Rome. Chiuderi and Ademollo also went abroad, the former to New York in 1968 and the latter to Harvard in 1969. Longhi and Celeghini stayed in Florence with a large group of young people who had just graduated or were about to.

	As already observed in the case of the 1930s for the school of Bruno Rossi and his colleagues, also in the case of the Gatto school, for various reasons, a group that had achieved world renown had disbanded and the task of rebuilding  the theory group in Florence took many years.

\section*{Acknowledgements}
The authors would like to thank Luisa Bonolis, Pier Andrea Mand\`o and Paolo Rossi for their useful and stimulating suggestions and for their help in reconstructing certain historical and scientific events.
\bibliography{sample1}

\begin{thebibliography}{109}
\providecommand{\natexlab}[1]{#1}
\providecommand{\url}[1]{\texttt{#1}}
\expandafter\ifx\csname urlstyle\endcsname\relax
  \providecommand{\doi}[1]{doi: #1}\else
  \providecommand{\doi}{doi: \begingroup \urlstyle{rm}\Url}\fi

\bibitem[Abetti(1933)]{abetti1933}
G.~Abetti.
\newblock {{Antonio Garbasso}}, 1933.
\newblock in Annuario della Regia Universit\`a di Firenze.

\bibitem[Ademollo and Gatto(1963)]{Ademollo1963MethodsFD}
M.~Ademollo and E.~Gatto.
\newblock {Methods for determining the spin of {$\Xi^-$}}.
\newblock \emph{Il Nuovo Cimento (1955-1965)}, 30:\penalty0 429--442, 1963.

\bibitem[Ademollo and Gatto(1964{\natexlab{a}})]{Ademollo1964CompleteST}
M.~Ademollo and R.~Gatto.
\newblock {Complete Spin Tests for Fermions}.
\newblock \emph{Physical Review}, 133:\penalty0 531--541, 1964{\natexlab{a}}.

\bibitem[Ademollo and
  Gatto(1964{\natexlab{b}})]{Ademollo1964NonrenormalizationTF}
M.~Ademollo and R.~Gatto.
\newblock {Nonrenormalization Theorem for the Strangeness-Violating Vector
  Currents}.
\newblock \emph{Physical Review Letters}, 13:\penalty0 264--266,
  1964{\natexlab{b}}.

\bibitem[Ademollo et~al.(1960)Ademollo, Bocciolini, di~Caporiacco, and
  Mand{\`o}]{Ademollo1960OnTI}
M.~Ademollo, M.~Bocciolini, G.~di~Caporiacco, and M.~Mand{\`o}.
\newblock {On the $^{196}Au^{m}$ isomeric state}.
\newblock \emph{Il Nuovo Cimento (1955-1965)}, 16:\penalty0 378--381, 1960.

\bibitem[Ademollo et~al.(1964)Ademollo, Gatto, and
  Preparata]{Ademollo1964TESTSFS}
M.~Ademollo, R.~Gatto, and G.~Preparata.
\newblock Tests for spin and parity of the b meson.
\newblock \emph{Physical Review Letters}, 12:\penalty0 462--465, 1964.

\bibitem[Ademollo et~al.(1967)Ademollo, Gatto, Longhi, and
  Veneziano]{Ademollo1967MixingSF}
M.~Ademollo, R.~Gatto, G.~Longhi, and G.~Veneziano.
\newblock Mixing scheme for chiral and collinear algebras.
\newblock \emph{Il Nuovo Cimento A (1965-1970)}, 47:\penalty0 334--338, 1967.

\bibitem[Ademollo et~al.(1969)Ademollo, Gatto, and Longhi]{Ademollo:1969wn}
M.~Ademollo, R.~Gatto, and G.~Longhi.
\newblock {Anomalous magnetic moments of nucleons and sidewise dispersion
  relations}.
\newblock \emph{Phys. Rev.}, 179:\penalty0 1601--1608, 1969.
\newblock \doi{10.1103/PhysRev.179.1601}.

\bibitem[Altarelli and Buccella(1964)]{Altarelli1964SinglePE}
G.~Altarelli and F.~Buccella.
\newblock Single photon emission in high-energy $e^+e^-$ collisions.
\newblock \emph{Il Nuovo Cimento (1955-1965)}, 34:\penalty0 1337--1346, 1964.

\bibitem[Altarelli et~al.(1965{\natexlab{a}})Altarelli, Buccella, and
  Gatto]{Altarelli1965SU6A}
G.~Altarelli, F.~Buccella, and R.~Gatto.
\newblock {$SU (6)$ and non leptonic hyperon decays}.
\newblock \emph{Physics Letters}, 1965{\natexlab{a}}.

\bibitem[Altarelli et~al.(1965{\natexlab{b}})Altarelli, Buccella, Preparata,
  and Gatto]{Altarelli1965TheVA}
G.~Altarelli, F.~Buccella, G.~Preparata, and R.~Gatto.
\newblock {The vector and axial couplings of broken SU(6)}.
\newblock \emph{Physics Letters}, 16:\penalty0 174--176, 1965{\natexlab{b}}.

\bibitem[Altarelli et~al.(1967)Altarelli, de~Gennaro, Celeghini, Longhi, and
  Gatto]{Altarelli1967TheoreticalCF}
G.~Altarelli, S.~de~Gennaro, E.~Celeghini, G.~Longhi, and R.~Gatto.
\newblock Theoretical calculations for electron-positron colliding-beam
  reactions.
\newblock \emph{Il Nuovo Cimento A (1965-1970)}, 47:\penalty0 113--135, 1967.

\bibitem[Armenise et~al.(1967)Armenise, Ghidini, Picciarelli, Romano, Forino,
  Gessaroli, Lendinara, Quareni, Quareni-Vignudelli, Cartacci, Dagliana,
  di~Caporiacco, Parrini, Barrier, Goussu, Laberrigue-Frolow, Mettel, Khanh,
  and Quinquard]{Armenise1967Ao2PI}
N.~Armenise, B.~Ghidini, V.~Picciarelli, A.~Romano, A.~Forino, R.~Gessaroli,
  L.~Lendinara, G.~Quareni, A.~Quareni-Vignudelli, A.~M. Cartacci, M.~G.
  Dagliana, G.~di~Caporiacco, G.~Parrini, M.~Barrier, O.~Goussu,
  J.~Laberrigue-Frolow, D.~Mettel, N.~H. Khanh, and J.~Quinquard.
\newblock $a_0^2$ production in $\pi^+$d interaction.
\newblock \emph{Physics Letters B}, 25:\penalty0 53--56, 1967.

\bibitem[Armenise et~al.(1968)Armenise, Ghidini, Picciarelli, Romano, Forino,
  Gessaroli, Lendinara, Quareni, Quareni-Vignudelli, Cartacci, Dagliana,
  di~Caporiacco, Parrini, Barrier, Laberrigue-Frolow, and
  Quinquard]{Armenise1968RI}
N.~Armenise, B.~Ghidini, V.~Picciarelli, A.~Romano, A.~Forino, R.~Gessaroli,
  L.~Lendinara, G.~Quareni, A.~Quareni-Vignudelli, A.~M. Cartacci, M.~G.
  Dagliana, G.~di~Caporiacco, G.~Parrini, M.~Barrier, J.~Laberrigue-Frolow, and
  J.~Quinquard.
\newblock $\rho\pi$ resonances in $\pi^+\pi^-\pi^0$ system.
\newblock \emph{Physics Letters B}, 26:\penalty0 336--340, 1968.

\bibitem[Armenise et~al.(1969)Armenise, Ghidini, Picciarelli, Romano,
  Silvestri, Cartacci, Dagliana, Caporiacco, Forino, Gessaroli, Lendinara, and
  Quareni-Vignudelli]{Armenise1969A3PI}
N.~Armenise, B.~Ghidini, V.~Picciarelli, A.~Romano, A.~Silvestri, A.~M.
  Cartacci, M.~G. Dagliana, G.~D. Caporiacco, A.~Forino, R.~Gessaroli,
  L.~Lendinara, and A.~Quareni-Vignudelli.
\newblock {$A_3^+$ production in $\pi$+d interactions at 5.1 GeV/c}.
\newblock \emph{Lettere al Nuovo Cimento (1969-1970)}, 2:\penalty0 501--509,
  1969.

\bibitem[Armenise et~al.(1970)Armenise, Ghidini, Picciarelli, Romano,
  Silvestri, Forino, Gessaroli, Lendinara, Quareni-Vignudelli, Cartacci,
  Dagliana, di~Caporiacco, Barrier, Mettel, and
  Quinquard]{Armenise1970QuasitwobodyPI}
N.~Armenise, B.~Ghidini, V.~Picciarelli, A.~Romano, A.~Silvestri, A.~Forino,
  R.~Gessaroli, L.~Lendinara, A.~Quareni-Vignudelli, A.~M. Cartacci, M.~G.
  Dagliana, G.~di~Caporiacco, M.~Barrier, D.~Mettel, and J.~Quinquard.
\newblock {Quasi-two-body processes in $\pi^+$-d inelastic reactions at 5.1
  GeV/c}.
\newblock \emph{Il Nuovo Cimento A (1965-1970)}, 65:\penalty0 637--653, 1970.

\bibitem[Ballario et~al.(1941)Ballario, Corte, and Prosperi]{bdcp1939}
C.~Ballario, M.~D. Corte, and M.~Prosperi.
\newblock Sulla componente dura e molle della radiazione cosmica fino a 575 m
  di acqua equivalente.
\newblock \emph{Rend. Acc. Reale d'Italia}, II:\penalty0 850, 1941.

\bibitem[Battimelli et~al.(2019)Battimelli, Buccella, and
  Napolitano]{battimelli}
G.~Battimelli, F.~Buccella, and P.~Napolitano.
\newblock {{Raoul Gatto, a great Italian scientist and teacher in theoretical
  elementary particle physics}}.
\newblock \emph{Quaderni di Storia della Fisica}, 1:\penalty0 145--169, 2019.
\newblock \doi{10.1393/qsf/i2019-10065-7}.

\bibitem[Benvenuti et~al.(1966)Benvenuti, Blasi, Maurenzig, and Sona]{bbms1966}
A.~Benvenuti, P.~Blasi, P.~Maurenzig, and P.~Sona.
\newblock On the backscattering of electrons on silicon detectors, 1966.
\newblock INFN-BE 66/4 Mars 1966.

\bibitem[Benvenuti et~al.(1965)Benvenuti, Blasi, Maurenzig, and
  Sona]{Benvenuti1965AVV}
A.~C. Benvenuti, P.~Blasi, P.~Maurenzig, and P.~Sona.
\newblock A versatile vacuum chamber for semiconductor beta and gamma
  detectors.
\newblock \emph{Nuclear Instruments and Methods}, 37:\penalty0 168--170, 1965.

\bibitem[Bizzeti and Corte(1959)]{Bizzeti1959OnTT}
P.~G. Bizzeti and M.~D. Corte.
\newblock On the thinning down of tracks of heavy nuclei in nuclear emulsions.
\newblock \emph{Il Nuovo Cimento (1955-1965)}, 11:\penalty0 317--333, 1959.

\bibitem[Bizzeti et~al.(1958)Bizzeti, Dagliana, Corte, and
  Tocci]{Bizzeti1958PhotometricAO}
P.~G. Bizzeti, M.~G. Dagliana, M.~D. Corte, and L.~R. Tocci.
\newblock Photometric analysis of the tracks in the nuclear emulsions.
\newblock \emph{Il Nuovo Cimento (1955-1965)}, 10:\penalty0 388--392, 1958.

\bibitem[Bizzeti et~al.(1963)Bizzeti, Cartacci, Dagliana, Corte, Tocci,
  B{\"o}bel, Tomasini, and Marzari-Chiesa]{Bizzeti1963OnTI}
P.~G. Bizzeti, A.~M. Cartacci, M.~G. Dagliana, M.~D. Corte, L.~R. Tocci,
  G.~B{\"o}bel, G.~Tomasini, and A.~Marzari-Chiesa.
\newblock {On the interactions of 25 GeV protons in nuclear emulsions. I}.
\newblock \emph{Il Nuovo Cimento (1955-1965)}, 27:\penalty0 6--13, 1963.

\bibitem[Bizzeti et~al.(1965)Bizzeti, Sona, Bocciolini, di~Caporiacco, Fazzini,
  and Mand{\`o}]{Bizzeti1965ErsicsonsFI}
P.~G. Bizzeti, A.~M.~B. Sona, M.~Bocciolini, G.~di~Caporiacco, T.~F. Fazzini,
  and M.~Mand{\`o}.
\newblock {Ericson's fluctuations in the photodisintegration of $Si^{28}$}.
\newblock \emph{Nuclear Physics}, 63:\penalty0 161--172, 1965.

\bibitem[Blasi et~al.(1966)Blasi, Maurenzig, and Sona]{Blasi1966ASF}
P.~Blasi, P.~Maurenzig, and P.~Sona.
\newblock A system for inserting radioactive sources in vacuum.
\newblock \emph{Nuclear Instruments and Methods}, 42:\penalty0 305--306, 1966.

\bibitem[Bonetti and Mazzoni(2006)]{bm2006}
A.~Bonetti and M.~Mazzoni.
\newblock {The Arcetri School of Physics}.
\newblock In P.~Redondi, G.~Sironi, P.~Tucci, and G.~Vegni, editors, \emph{The
  Scientific Legacy of Beppo Occhialini}, pages 3--34. SIF-Springer, Bologna
  and Heidelberg, 2006.

\bibitem[Bonetti and Mazzoni(2007)]{bm2007}
A.~Bonetti and M.~Mazzoni, editors.
\newblock \emph{{L'Universit\`a di Firenze nel centenario della nascita di
  Giuseppe Occhialini (1907-1993})}.
\newblock Firenze University Press., Firenze, 2007.

\bibitem[Bonolis(2011)]{Bonolis:2011ph}
L.~Bonolis.
\newblock {Walther Bothe and Bruno Rossi: the birth and development of
  coincidence methods in cosmic-ray physics}.
\newblock \emph{Am. J. Phys.}, 79:\penalty0 1133, 2011.
\newblock \doi{10.1119/1.3619808}.

\bibitem[Bonolis(2014)]{Bonolis:2012dm}
L.~Bonolis.
\newblock {From cosmic ray physics to cosmic ray astronomy: Bruno Rossi and the
  opening of new windows on the universe}.
\newblock \emph{Astropart. Phys.}, 53:\penalty0 67--85, 2014.
\newblock \doi{10.1016/j.astropartphys.2013.05.008}.

\bibitem[Borchi(1965)]{Borchi1965RadiativeMC}
E.~Borchi.
\newblock Radiative muon capture from nuclei.
\newblock \emph{Il Nuovo Cimento (1955-1965)}, 35:\penalty0 1077--1104, 1965.

\bibitem[Borchi and Gatto(1965)]{Borchi1965AssignmentOH}
E.~Borchi and R.~Gatto.
\newblock {Assignment of higher boson resonances to a $p$-wave multiplet of
  $SU(6)$ }.
\newblock \emph{Physics Letters}, 14:\penalty0 352--354, 1965.

\bibitem[Brunetti(1933)]{brunetti1933}
R.~Brunetti.
\newblock {Antonio Garbasso, La vita, il pensiero e l'opera scientifica}.
\newblock \emph{Il Nuovo Cimento}, 10:\penalty0 129--152, 1933.

\bibitem[Buccella and Gatto(1965)]{Buccella1965TheCG}
F.~Buccella and R.~Gatto.
\newblock {The collinear groups $(SU(3)\otimes SU(3))_{coll}$ and $SU(2)\otimes
  SU(2) \otimes W(\sigma_zY))_{coll}$}.
\newblock \emph{Il Nuovo Cimento A (1965-1970)}, 40:\penalty0 684--689, 1965.

\bibitem[Buccella et~al.(1966)Buccella, Veneziano, Gatto, and
  Okubo]{Buccella1966NecessityOA}
F.~Buccella, G.~Veneziano, R.~Gatto, and S.~Okubo.
\newblock {Necessity of Additional Unitary-Antisymmetric q -Number Terms in the
  Commutators of Spatial Current Components}.
\newblock \emph{Physical Review}, 149:\penalty0 1268--1272, 1966.

\bibitem[Cabibbo and Gatto(1961{\natexlab{a}})]{Cabibbo:1961sz}
N.~Cabibbo and R.~Gatto.
\newblock {{Electron Positron Colliding Beam Experiments}}.
\newblock \emph{Phys. Rev.}, 124:\penalty0 1577--1595, 1961{\natexlab{a}}.
\newblock \doi{10.1103/PhysRev.124.1577}.

\bibitem[Cabibbo and Gatto(1961{\natexlab{b}})]{bibbia}
N.~Cabibbo and R.~Gatto.
\newblock {{Theoretical Discussion of Possible Experiments with Electron
  Positron Colliding Beams}}.
\newblock \emph{Il Nuovo Cimento}, 20:\penalty0 185--193, 1961{\natexlab{b}}.
\newblock \doi{10.1103/PhysRev.124.1577}.

\bibitem[Cartacci(2014)]{Cartacci2014}
A.~Cartacci.
\newblock {The plates group of the {Antonio Garbasso Institute of Florence}
  (1953-1983)}.
\newblock \emph{Il Colle di Galileo}, 3\penalty0 (1):\penalty0 7--14, 2014.

\bibitem[Casalbuoni and Dominici(2018)]{Casalbuoni_Dominici_2018}
R.~Casalbuoni and D.~Dominici.
\newblock The teacher of the gattini (kittens).
\newblock \emph{Il Colle di Galileo}, 7-2:\penalty0 47--69, 2018.

\bibitem[Casalbuoni et~al.(1969)Casalbuoni, Longhi, and
  Gatto]{Casalbuoni:1969gv}
R.~Casalbuoni, G.~Longhi, and R.~Gatto.
\newblock {The vector and axial $n-n^*$ vertices. Calculation from the ward
  identity and the sidewise dispersion relations}.
\newblock \emph{Nuovo Cim. A}, 62:\penalty0 319--331, 1969.
\newblock \doi{10.1007/BF02731812}.

\bibitem[Casalbuoni et~al.(2019)Casalbuoni, Dominici, and Pelosi]{cdp2019}
R.~Casalbuoni, D.~Dominici, and G.~Pelosi.
\newblock \emph{{Enrico Fermi a Firenze. Le Lezioni di Meccanica Razionale al
  biennio propedeutico agli studi di Ingegneria: 1924-1926}}.
\newblock Florence University Press, Firenze, 2019.

\bibitem[Casalbuoni et~al.(2021)Casalbuoni, Dominici, and Mazzoni]{cdm2021}
R.~Casalbuoni, D.~Dominici, and M.~Mazzoni.
\newblock \emph{{Lo spirito di Arcetri. A cento anni dalla nascita
  dell'Istituto di Fisica dell'Universit\`a di Firenze}}.
\newblock Florence University Press, Firenze, 2021.

\bibitem[Celeghini and Gatto(1964{\natexlab{a}})]{Celeghini1964PossibleDO}
E.~Celeghini and R.~Gatto.
\newblock Possible determination of the $\eta_0$ lifetime with
  electron-positron colliding beams.
\newblock \emph{Il Nuovo Cimento (1955-1965)}, 33:\penalty0 657--662,
  1964{\natexlab{a}}.

\bibitem[Celeghini and Gatto(1964{\natexlab{b}})]{Celeghini1964PossibleMT}
E.~Celeghini and R.~Gatto.
\newblock Possible method to determine the $\eta_0$ lifetime.
\newblock \emph{Physics Letters}, 10:\penalty0 245--247, 1964{\natexlab{b}}.

\bibitem[Celeghini and Gatto(1966)]{Celeghini1966LocalCR}
E.~Celeghini and R.~Gatto.
\newblock Local commutation relations of vector and axial currents and their
  approximate saturation.
\newblock \emph{Il Nuovo Cimento A (1971-1996)}, 43:\penalty0 219--224, 1966.

\bibitem[Chiuderi and Morpurgo(1961)]{Chiuderi1961CoulombPO}
C.~Chiuderi and G.~Morpurgo.
\newblock Coulomb photoproduction of $\pi_0$ at high energy and $\pi_0$
  lifetime.
\newblock \emph{Il Nuovo Cimento (1955-1965)}, 19:\penalty0 497--511, 1961.

\bibitem[Cordella and Sebastiani(2000)]{cordella2000}
F.~Cordella and F.~Sebastiani.
\newblock {Sul percorso di Fermi verso la statistica quantistica}.
\newblock \emph{{Il Nuovo Saggiatore}}, 16:\penalty0 11--22, 2000.

\bibitem[Cordella et~al.(2001)Cordella, De~Gregorio, and
  Sebastiani]{cordella2001}
F.~Cordella, A.~De~Gregorio, and F.~Sebastiani.
\newblock \emph{{{Enrico Fermi. Gli anni italiani}}}.
\newblock Editori Riuniti, Roma, 2001.

\bibitem[Della~Corte(2001)]{dcl}
L.~Della~Corte.
\newblock {Commemorazione di Michele Della Corte. Firenze 21 settembre 1999}.
\newblock Unpublished paper, Fondo Della Corte, Biblioteca di Scienze,
  Universit\`a di Firenze, 2001.

\bibitem[Della~Corte(1954)]{Corte1954}
M.~Della~Corte.
\newblock The grain density and the process of track formation in nuclear
  emulsions.
\newblock \emph{Il Nuovo Cimento (1943-1954)}, 12:\penalty0 28--36, 1954.

\bibitem[Della~Corte(1956)]{dc1956}
M.~Della~Corte.
\newblock {Analisi fotometrica delle tracce nelle emulsioni nucleari I.
  Dispositivo sperimentale}.
\newblock \emph{Nuovo Cimento}, 4:\penalty0 1565--1569, 1956.

\bibitem[Della~Corte(1991)]{dc99}
M.~Della~Corte.
\newblock Ai miei nipoti.
\newblock Unpublished paper, Fondo Della Corte, Biblioteca di Scienze,
  Universit\`a di Firenze, 1991.

\bibitem[Della~Corte and Giovannozzi(1951)]{Corte1951AnIA}
M.~Della~Corte and M.~Giovannozzi.
\newblock An intrinsic angular analysis of cosmic ray stars.
\newblock \emph{Il Nuovo Cimento (1943-1954)}, 8:\penalty0 741--748, 1951.

\bibitem[Della~Corte and Ramat(1952)]{DellaCorte1952PhotometricMO}
M.~Della~Corte and M.~Ramat.
\newblock {Photometric Measurements of Tracks in Nuclear Emulsions}.
\newblock \emph{Il Nuovo Cimento (1943-1954)}, 9:\penalty0 605--609, 1952.

\bibitem[Della~Corte et~al.(1947)Della~Corte, Fazzini, and
  Franchetti]{Corte1947TheMS}
M.~Della~Corte, T.~F. Fazzini, and S.~Franchetti.
\newblock {The Meson Spectrum Near Sea-Level}.
\newblock \emph{Nature}, 159:\penalty0 845, 1947.

\bibitem[Della~Corte et~al.(1953)Della~Corte, Ramat, and
  Ronchi]{Corte1953TheGD}
M.~Della~Corte, M.~Ramat, and L.~A. Ronchi.
\newblock The grain density and the process of track formation in nuclear
  emulsions.
\newblock \emph{Il Nuovo Cimento (1943-1954)}, 10:\penalty0 958--970, 1953.

\bibitem[Della~Corte et~al.(1955)Della~Corte, Fazzini, and
  Sona]{Corte1955AnelasticSO}
M.~Della~Corte, T.~F. Fazzini, and A.~M. Sona.
\newblock Anelastic scattering of $\pi$-mesons on carbon.
\newblock \emph{Il Nuovo Cimento (1955-1965)}, 2:\penalty0 1345--1349, 1955.

\bibitem[di~Caporiacco and Giovannozzi(1956)]{Caporiacco1956CloudCS}
G.~di~Caporiacco and M.~Giovannozzi.
\newblock {Cloud chamber study of cosmic ray electronic showers under dense
  materials (II)}.
\newblock \emph{Il Nuovo Cimento (1955-1965)}, 3:\penalty0 305--317, 1956.

\bibitem[Fermi(1926{\natexlab{a}})]{fermistat1}
E.~Fermi.
\newblock Sulla quantizzazione del gas perfetto monoatomico.
\newblock \emph{{Rendiconti Lincei}}, 3:\penalty0 145--149, 1926{\natexlab{a}}.

\bibitem[Fermi(1926{\natexlab{b}})]{fermistat2}
E.~Fermi.
\newblock Zur quantelung des idealen einatomigen gases.
\newblock \emph{{Z. f\"ur Physik}}, 36:\penalty0 902--912, 1926{\natexlab{b}}.

\bibitem[Fermi(1927)]{fermi1927}
E.~Fermi.
\newblock \emph{{Atti R. Accademia dei Lincei Rend. }}, 6:\penalty0 602, 1927.

\bibitem[Fermi(1962)]{fermioc}
E.~Fermi.
\newblock \emph{{{Note e Memorie (Collected papers)}}}.
\newblock Accademia Nazionale dei Lincei, Roma, 1962.

\bibitem[Fermi and Rasetti(1925{\natexlab{a}})]{rasfermi1925}
E.~Fermi and F.~Rasetti.
\newblock Effect of an alternating magnetic field on the polarization of the
  resonance radiation of mercury vapour.
\newblock \emph{{Nature}}, 115:\penalty0 764, 1925{\natexlab{a}}.

\bibitem[Fermi and Rasetti(1925{\natexlab{b}})]{rasfermi1925a}
E.~Fermi and F.~Rasetti.
\newblock {\"Uber den Einfluss eines wechselnden magnetischen Feldes auf die
  Polarisation}der resonanzstrahlung.
\newblock \emph{{Z. f\"ur Physik}}, 33:\penalty0 246--250, 1925{\natexlab{b}}.

\bibitem[Fermi and Rossi(1933)]{rossifermi}
E.~Fermi and B.~Rossi.
\newblock {Azione del campo magnetico terrestre sulla radiazione penetrante}.
\newblock \emph{Atti R. Accademia dei Lincei Rend.}, 17:\penalty0 346--350,
  1933.

\bibitem[Fermi(1954)]{caponl}
L.~Fermi.
\newblock \emph{{{Atoms in the family. My life with Enrico Fermi}}}.
\newblock University of Chicago press, Chicago, 1954.

\bibitem[Ferretti et~al.(1957)Ferretti, Quareni, Corte, and
  Fazzini]{Ferretti1957ProtonSA}
L.~Ferretti, G.~Quareni, M.~D. Corte, and T.~F. Fazzini.
\newblock {$\pi^+$-proton scattering at 83 MeV}.
\newblock \emph{Il Nuovo Cimento (1955-1965)}, 5:\penalty0 1660--1662, 1957.

\bibitem[Fidecaro(2017)]{Fidecaro:2017aee}
G.~Fidecaro.
\newblock {Discovery of the ${\pi} \to e {\nu}$ Decay: Rare and Precious}.
\newblock \emph{{Adv. Ser. Direct. High Energy Phys.}}, 27:\penalty0 29--32,
  2017.

\bibitem[Fowler(1926)]{fowler1926}
R.~Fowler.
\newblock \emph{{Proceedings of the Royal Society}}, A87:\penalty0 114, 1926.

\bibitem[Franchetti(1954)]{Franchetti1954RleOE}
S.~Franchetti.
\newblock {R{o}le of exchange forces in the problem of helium II}.
\newblock \emph{Il Nuovo Cimento (1943-1954)}, 12:\penalty0 743--768, 1954.

\bibitem[Franchetti(1955{\natexlab{a}})]{Franchetti1955ProblemsOF}
S.~Franchetti.
\newblock {Problems of film formation and flow in liquid helium II}.
\newblock \emph{Il Nuovo Cimento (1955-1965)}, 2:\penalty0 1127--1129,
  1955{\natexlab{a}}.

\bibitem[Franchetti(1955{\natexlab{b}})]{Franchetti1955TentativeOT}
S.~Franchetti.
\newblock {Tentative on the r{o}le of exchange forces in the problem of He II}.
\newblock \emph{Il Nuovo Cimento (1955-1965)}, 1:\penalty0 159--160,
  1955{\natexlab{b}}.

\bibitem[Franchetti(1960)]{Franchetti1960SomeRO}
S.~Franchetti.
\newblock Some remarks on the theory of the liquid helium film.
\newblock \emph{Il Nuovo Cimento (1955-1965)}, 16:\penalty0 1158--1159, 1960.

\bibitem[Franchetti(1961)]{Franchetti1961OnTS}
S.~Franchetti.
\newblock On the structure of liquid {4He} from the elastic scattering of
  neutrons.
\newblock \emph{Il Nuovo Cimento (1955-1965)}, 22:\penalty0 374--394, 1961.

\bibitem[Franchetti(1964)]{Franchetti1964OnTF}
S.~Franchetti.
\newblock On the flowing helium film.
\newblock \emph{Il Nuovo Cimento (1955-1965)}, 33:\penalty0 144--147, 1964.

\bibitem[Gasperini et~al.(2004)Gasperini, Mazzoni, and Righini]{gmr2004}
A.~Gasperini, M.~Mazzoni, and A.~Righini.
\newblock {{L'evoluzione della Torre Solare di Arcetri nel carteggio
  Hale-Abetti}}.
\newblock \emph{Giornale di Astronomia}, 3:\penalty0 23--31, 2004.

\bibitem[Gatto et~al.(1964)Gatto, Longhi, and Furlan]{fgl1964}
R.~Gatto, G.~Longhi, and G.~Furlan.
\newblock Radiative corrections to $e^+e^- \to \mu^+ \mu^-$.
\newblock \emph{Phys. Lett}, 12:\penalty0 262, 1964.

\bibitem[Goodstein(1982)]{goods}
J.~Goodstein.
\newblock {Franco Rasetti (1901-2001)}, 1982.
\newblock Interview, February 4, Archives California Institute of Technology
  \url{http://oralhistories.library.caltech.edu/70/1/OH\_Rasetti.pdf}.

\bibitem[Grandolfo et~al.(2017)Grandolfo, Napolitano, Risica, and
  Tabet]{sanita}
M.~Grandolfo, N.~Napolitano, S.~Risica, and E.~Tabet.
\newblock \emph{{{Il Laboratorio di Fisica dell'Istituto Superiore di Sanit\`a
  }}}.
\newblock Istituto Superiore di Sanit\`a, Roma, 2017.

\bibitem[Guarnieri(2019)]{guarnieri}
P.~Guarnieri.
\newblock \emph{L' emigrazione intellettuale dall'Italia fascista}.
\newblock Florence University Press, Firenze, 2019.

\bibitem[Guerra and Robotti(2015)]{gr2015}
F.~Guerra and N.~Robotti.
\newblock \emph{{{Enrico Fermi e il quaderno ritrovato 20 marzo 1934 La vera
  storia della scoperta della radioattivit\`a indotta da neutroni}}}.
\newblock SIF, Bologna, 2015.

\bibitem[Iurato and Rossi(2019)]{iur2019}
G.~Iurato and P.~Rossi.
\newblock \emph{La scuola pisana di fisica (1840-1950)}.
\newblock Pisa University Press, Pisa, 2019.

\bibitem[La~Rana and Rossi(2020)]{larana1}
A.~La~Rana and P.~Rossi.
\newblock {The blossoming of quantum mechanics in Italy: the roots, the context
  and the first spreading in Italian universities (1900-1947)}.
\newblock \emph{{Eur. Phys. J. H}}, 45:\penalty0 237--252, 2020.

\bibitem[Maiani(2008)]{maiani}
L.~Maiani.
\newblock \emph{Fisico: andare a caccia di particelle}.
\newblock Zanichelli, Bologna, 2008.

\bibitem[Mand\`o(1951)]{mandFG1es}
M.~Mand\`o.
\newblock \emph{{{Esercizi e problemi di Fisica. I. Meccanica, Termologia}}}.
\newblock Libreria Universitaria, Bologna, 1951.

\bibitem[Mand\`o(1961)]{mandFG1}
M.~Mand\`o.
\newblock \emph{{Lezioni di Fisica Generale I}}.
\newblock Libreria Universitaria, Bologna, 1961.

\bibitem[Mand{\`o}(1962)]{Mand1962The7LipA}
M.~Mand{\`o}.
\newblock {The $^7Li+p$ $\gamma$-radiation as a tool for the detection of
  nuclear cross-section fluctuations}.
\newblock \emph{Il Nuovo Cimento (1955-1965)}, 26:\penalty0 1416--1418, 1962.

\bibitem[Mand\`o(1986)]{mandstoria}
M.~Mand\`o.
\newblock Notizie sugli studi di fisica (1859-1949).
\newblock In \emph{{Storia dell'Ateneo fiorentino}}, volume~1, Firenze, 1986.
  Parretti Grafiche.

\bibitem[Mand{\`o} and Ronchi(1952)]{Mand1952OnTE}
M.~Mand{\`o} and L.~A. Ronchi.
\newblock On the energy range relation for fast muons in rock.
\newblock \emph{Il Nuovo Cimento (1943-1954)}, 9:\penalty0 517--529, 1952.

\bibitem[Mand{\`o} and Sona(1953)]{ms1953}
M.~Mand{\`o} and P.~Sona.
\newblock Sull'applicabilit\`a del concetto di percorso per l'assorbimento dei
  muoni sotto terra.
\newblock \emph{Il Nuovo Cimento}, 10:\penalty0 1275--1287, 1953.

\bibitem[Mand\`o(2014)]{mand2014}
P.~Mand\`o.
\newblock {La fisica nucleare applicata negli ultimi anni ad Arcetri e la
  nascita di nuove attivit\`a al LABEC del Polo Scientifico a Sesto
  Fiorentino}.
\newblock \emph{{Il Colle di Galileo}}, 3-1:\penalty0 15--29, 2014.

\bibitem[Marzari-Chiesa et~al.(1963)Marzari-Chiesa, Rinaudo, Ciurlo, Picasso,
  and Cartacci]{MarzariChiesa1963OnTI}
A.~Marzari-Chiesa, G.~Rinaudo, S.~Ciurlo, E.~Picasso, and A.~M. Cartacci.
\newblock {On the interactions of 25 GeV protons in nuclear emulsions. II}.
\newblock \emph{Il Nuovo Cimento (1955-1965)}, 27:\penalty0 155--163, 1963.

\bibitem[Pais(1988)]{pais1985}
A.~Pais.
\newblock \emph{{Inward Bound: Of Matter and Forces in the Physical World}}.
\newblock Oxford University Press, New York, 1988.

\bibitem[Palla(1978)]{palla}
F.~Palla.
\newblock \emph{{Firenze nel regime fascista (1923-1934)}}.
\newblock Olschki, Firenze, 1978.

\bibitem[Persico(1950)]{persico1950}
E.~Persico.
\newblock \emph{{{Fundamentals of Quantum Mechanics}}}.
\newblock Prentice-Hall, Englewood Cliffs, NJ, 1950.

\bibitem[Peruzzi and Talas(2007)]{Peruzzi:2007zz}
G.~Peruzzi and S.~Talas.
\newblock {The Italian contributions to cosmic-ray physics from Bruno Rossi to
  the G-stack. A new window into the inexhaustible wealth of nature}.
\newblock \emph{Riv. Nuovo Cim.}, 30\penalty0 (5):\penalty0 197--257, 2007.
\newblock \doi{10.1393/ncr/i2007-10020-0}.

\bibitem[Pestre(1984)]{pestre1984}
D.~Pestre.
\newblock {Prehistory of CERN: the first suggestions (1949-June 1950)}, 1984.
\newblock CHS-3 March 1984.

\bibitem[Pontecorvo(1993)]{pontecorvo1993}
B.~Pontecorvo.
\newblock \emph{{{Enrico Fermi}}}.
\newblock Studio Tesi, Pordenone, 1993.

\bibitem[Preparata(2002)]{preparata}
G.~Preparata.
\newblock \emph{Dai quark ai cristalli}.
\newblock Bollati Boringhieri, Torino, 2002.

\bibitem[Quareni et~al.(1965)Quareni, Vignudelli, Cartacci, Dagliana, della
  Corte, Tomasini, Gainotti, Lamborizio, Mora, and Ortalli]{Quareni1965TheRO}
G.~Quareni, A.~Q. Vignudelli, A.~M. Cartacci, M.~G. Dagliana, M.~della Corte,
  G.~Tomasini, A.~Gainotti, C.~Lamborizio, S.~Mora, and I.~Ortalli.
\newblock {The rate of the radiative decay $\Sigma^+\to p+\gamma$}.
\newblock \emph{Il Nuovo Cimento A (1965-1970)}, 40:\penalty0 928--934, 1965.

\bibitem[Ricci and Maurenzig(1969)]{Ricci1969The1P}
R.~A. Ricci and P.~Maurenzig.
\newblock The 1f7/2 problem in nuclear spectroscopy.
\newblock \emph{La Rivista del Nuovo Cimento}, 1:\penalty0 291--354, 1969.

\bibitem[Rossi(1930{\natexlab{a}})]{RossiMethodOR}
B.~Rossi.
\newblock {Method of Registering Multiple Simultaneous Impulses of Several
  Geiger's Counters}.
\newblock \emph{Nature}, 125:\penalty0 636, 1930{\natexlab{a}}.

\bibitem[Rossi(1930{\natexlab{b}})]{RossiOnTM}
B.~Rossi.
\newblock {On the Magnetic Deflection of Cosmic Rays}.
\newblock \emph{Physical Review}, 36:\penalty0 606, 1930{\natexlab{b}}.

\bibitem[Rossi(1931)]{Rossi1931MagneticEO}
B.~Rossi.
\newblock {Magnetic Experiments on the Cosmic Rays}.
\newblock \emph{Nature}, 128:\penalty0 300--301, 1931.

\bibitem[Rossi(1932)]{Rossitriple}
B.~Rossi.
\newblock {Nachweis einer Sekund\"arstrahlung der durchdringenden
  Korpuskularstrahlung}.
\newblock \emph{Physikalische Zeitschrift}, 33:\penalty0 30, 1932.

\bibitem[Rossi(1987)]{rossibook}
B.~Rossi.
\newblock \emph{{Momenti nella vita di uno scienziato}}.
\newblock Zanichelli, Bologna, 1987.

\bibitem[Sommerfeld(1927)]{sommerfeld1927}
A.~Sommerfeld.
\newblock \emph{{Naturwissenschaften}}, 15:\penalty0 824, 1927.

\bibitem[Taccetti(2017)]{Taccetti2017}
N.~Taccetti.
\newblock {Physics with accelerators at Arcetri. A short chronicle dedicated to
  Tito Fazzini who was one of its leading protagonists}.
\newblock \emph{Il Colle di Galileo}, 6\penalty0 (1):\penalty0 19--38, Apr.
  2017.

\bibitem[Thomas(1927)]{thomas1927}
L.~Thomas.
\newblock \emph{{Proc. Cambridge Phil. Soc.}}, 23:\penalty0 542, 1927.

\bibitem[Veneziano(1968)]{Veneziano:1968yb}
G.~Veneziano.
\newblock {{Construction of a crossing - symmetric, Regge behaved amplitude for
  linearly rising trajectories}}.
\newblock \emph{Nuovo Cim. A}, 57:\penalty0 190--197, 1968.
\newblock \doi{10.1007/BF02824451}.

\end{thebibliography}
\end{document}